\documentclass[11pt,a4paper]{article}
\usepackage[utf8]{inputenc}
\usepackage{a4wide}
\usepackage{amsmath,amssym  b}
\usepackage{amsfonts}
\usepackage{cite}
\usepackage{color}
\usepackage{adjustbox}
\usepackage{multirow}
\usepackage{pdflscape}
\usepackage{textcomp}
\usepackage{gensymb}
\usepackage{graphicx}
\usepackage{diagbox}
\usepackage{pifont}
\usepackage{bigstrut}
\usepackage[small,bf]{caption}
\usepackage{tabularx}
\usepackage{booktabs}
\usepackage{stackengine}
\usepackage{mathtools}
\usepackage{enumerate}
\usepackage{makecell}
\usepackage{listings}
\usepackage{bbm}
\usepackage[unicode]{hyperref} 
\usepackage{setspace}

\definecolor{greenLinks}{rgb}{0, 0.6, 0} 
\definecolor{blueLinks}{rgb}{0, 0, 0.6}
\definecolor{redLinks}{rgb}{0.6, 0, 0}
\definecolor{eprintLinks}{rgb}{0.4, 0.4, 0.4}
\definecolor{journalLinks}{rgb}{0.6, 0, 0}
\DeclareMathOperator{\diag}{diag}
\DeclareMathOperator{\sign}{sign}

\begin{document}

\pagenumbering{Alph}
\begin{titlepage}

\vspace*{15mm}

\begin{center}
{ \bf\LARGE {A $Z_4$ symmetric inverse seesaw model for neutrino \\[2mm] masses and FIMP dark matter}}\\[8mm]
Ziye Wang$^{\,a,}$\footnote{Email: \href{mailto:zyewang@163.com}{\texttt{zyewang@163.com}}},
Yakefu Reyimuaji$^{\,a,}$\footnote{Email: \href{mailto:yreyi@hotmail.com}{\texttt{yreyi@hotmail.com}}},
Nijiati Yalikun$^{\,a,}$\footnote{Email: \href{mailto:nijiati@xju.edu.cn}{\texttt{nijiati@xju.edu.cn}}}\\
\vspace{8mm}
$^{a}$\,{\it School of Physical Science and Technology, Xinjiang University, Urumqi 830017, China} \\

\vspace{2mm}

\end{center}
\setcounter{footnote}{0} 

\vspace{8mm}

\begin{abstract}
\noindent A theoretical framework based on a spontaneously broken $Z_4$ symmetry is proposed to simultaneously explain neutrino mass generation via the inverse seesaw mechanism and dark matter (DM) production through a freeze-in scenario. This work extends the standard model with right-handed neutrinos $N_i$, additional fermions $\chi_i$, and a complex scalar $S$. An unbroken $Z_2$ subgroup ensures the stability of the DM candidate, whose relic abundance is dominantly produced via decay and scattering processes involving heavy singlet fermions. Phenomenological analyses show that this relatively minimal construction accommodates the observed neutrino oscillation parameters, consistent with the latest global fit data. Furthermore, the model successfully reproduces the observed DM relic density within the parameter space relevant to neutrino phenomenology, establishing a connection between neutrino properties and DM production.

\end{abstract}

\end{titlepage}
\pagenumbering{arabic}



\section{Introduction}
\label{sec:intro}
The search for physics beyond the Standard Model (SM) requires an explanation of the origin of neutrino masses and the nature of DM. Neutrino oscillation experiments have firmly established that neutrinos are massive and undergo flavor mixing~\cite{Kajita:2016cak,McDonald:2016ixn}. The existence of DM is indicated indirectly through astronomical and cosmological observations~\cite{Bertone:2004pz,Arbey:2021gdg,Cirelli:2024ssz}. Yet, the underlying mechanisms explaining these phenomena remain unknown, motivating theoretical frameworks that can simultaneously explain neutrino masses and DM. 

Among the mechanisms proposed to explain neutrino masses, seesaw mechanisms are particularly compelling, as they naturally suppress neutrino masses through the introduction of new heavy states, whose masses are typically around $10^{14}$ GeV. The inverse seesaw mechanism offers a testable alternative, with mediators at the TeV scale, making it accessible to current and future collider experiments~\cite{Mohapatra:1986bd,Das:2012ze,Das:2019pua,CarcamoHernandez:2019eme,CentellesChulia:2020dfh}. This framework often extends the fermion sector, providing a natural setting for DM candidates when protected by appropriate symmetries~\cite{Law:2012mj,Abada:2014zra,Mukherjee:2015axj,Abdallah:2019svm,Pongkitivanichkul:2019cvm,Mandal:2019oth,Gu:2019gzy,Fernandez-Martinez:2021ypo,Abada:2021yot,Zhang:2021olk,Gogoi:2023jzl}.

The production of DM may proceed through various cosmological mechanisms, such as thermal freeze-out, freeze-in, gravitational production, asymmetric scenarios, the misalignment mechanism, and production via topological defects or cosmological phase transitions~\cite{Cirelli:2024ssz}. In contrast to the canonical freeze-out paradigm, where DM particles initially maintain thermal equilibrium with the SM plasma before decoupling as the universe expands~\cite{Zeldovich:1965gev,Scherrer:1985zt,Gondolo:1990dk,Steigman:2012nb,Schumann:2019eaa,Frumkin:2022ror, delaTorre:2023nfk}, the freeze-in mechanism generates the observed abundance through feeble, non-thermalizing interactions with SM particles~~\cite{Hall:2009bx,Elahi:2014fsa,Biswas:2016bfo,Bernal:2017kxu,Chianese:2018dsz,Becker:2018rve,Chianese:2019epo,Allahverdi:2019jsc,Cosme:2020mck,Das:2021nqj,Yaguna:2023kyu,Abada:2023mib,Xu:2023xva}. This framework is naturally realized by Feebly Interacting Massive Particles (FIMPs), whose extremely weak couplings allow them to evade stringent direct and indirect detection constraints. This work investigates the freeze-in mechanism as a viable pathway to generate the observed DM relic density.

Discrete symmetries provide a useful tool for constructing viable DM models while simultaneously addressing neutrino mass generation. In particular, $Z_N$ symmetries can stabilize DM candidates by forbidding their decay into SM particles~\cite{Schmaltz:2017oov,Yaguna:2019cvp,Yaguna:2021vhb,Hepburn:2022pin,Liu:2023kil,Kim:2024cwp}. In this paper, we explore a model based on a spontaneously broken $Z_4$ symmetry that preserves a $Z_2$ subgroup, ensuring the stability of DM. In this construction, the $Z_4$ symmetry naturally separates the transformation properties of SM fields from the new fermionic sector, preventing unwanted interactions while maintaining the stability of the lightest $Z_2$-odd particle.

The proposed model demonstrates how a minimal extension of the SM can simultaneously explain neutrino masses through the inverse seesaw mechanism and generate the correct DM abundance via freeze-in production. The heavy singlet fermions $N_i$ and $\chi_i$ mediate neutrino mass generation through their interactions with the scalar field $S$, while their decays and scatterings dominantly contribute to the freeze-in production of DM. This unified framework connects neutrino phenomenology with DM production in a testable, symmetry-protected scenario.

The structure of this paper is organized as follows. Section~\ref{sec:model} presents our theoretical framework, including field content, symmetry assignments, and the resulting mass matrices. Section~\ref{sec:pheno} contains our phenomenological analysis of both neutrino oscillations and DM production mechanisms. Section~\ref{sec:conclusions} summarizes our results and outlines future research directions. The technical details of the perturbative diagonalization of the heavy sector mass matrix are presented in the Appendix~\ref{app:srpertblock}.

\section{The model description}
\label{sec:model}

Here, we present details of a model that extends the SM based on a $Z_4$ symmetry. In this framework, all SM fields, except for the leptons, remain invariant under this symmetry. The model introduces new fields that are invariant under SM symmetries, including right-handed neutrinos $N_i$, additional right-handed fermion singlets $\chi_i$, and a complex scalar field $S$. These new fields are singlets under the SM gauge groups, but transform non-trivially under the $Z_4$ symmetry. The scalar field $S$ may acquire a nonzero vacuum expectation value (VEV), which consequently leads to the spontaneous breaking of the $Z_4$ symmetry to its subgroup $Z_2$. The field content and the corresponding transformations under the electroweak gauge symmetries of the SM, the broken $Z_4$ symmetry, and the unbroken $Z_2$, are listed in Table~\ref{tab:Z4chasign}. The charge assignments forbid direct Majorana mass terms for the right-handed neutrinos and preclude the standard type-I seesaw mechanism, ensuring the inverse seesaw mechanism dominates neutrino mass generation, while the unbroken $Z_2$ symmetry guarantees DM stability.

\begin{table}[!h]
\setstretch{1.3}
\centering
 \begin{tabularx}{0.9\textwidth} { 
  | >{\centering\arraybackslash}X 
  || >{\centering\arraybackslash}X
  | >{\centering\arraybackslash}X 
  | >{\centering\arraybackslash}X 
  | >{\centering\arraybackslash}X 
  || >{\centering\arraybackslash}X 
  | >{\centering\arraybackslash}X | }
			\hline 
			Fields & $L$ & $e_R$ & $N$ & $\chi$ &  $S$ & $H$ \\ 
            \hline
            $SU_\mathrm{L}$ & 2 & 1 & 1 & 1 & 1 & 2 \\
            \hline
            $U(1)_\mathrm{Y}$  & $-\frac{1}{2}$ & -1 & 0 & 0 & 0 & $\frac{1}{2}$\\
            \hline
            $Z_4\to Z_2$ & $z\to -1$ & $z\to -1$ & $z\to -1$ & $z^3\to -1$ & $z^2\to 1$ & $1\to 1$ \\ 
            \hline
\end{tabularx}
	\caption{Field content and symmetry transformations under the SM gauge group $SU(2)_{\rm L}\times U(1)_{\rm Y}$, the $Z_4$ symmetry (with generator $z=e^{i\pi/2}$), and the residual $Z_2$ symmetry. Other SM fields which are not shown here transform trivially under these discrete symmetries.}
	\label{tab:Z4chasign}
\end{table}

The Yukawa sector of the model is described by the Lagrangian
\begin{equation}
\mathcal{L} =Y_E\overline{L} e_R H +Y_\nu\overline{L}N\tilde{H}+M\overline{N^c}\chi+\lambda S\, \overline{\chi^c}\chi +\lambda' S\overline{N^c}N + \mathrm{h.c.}.
\label{eq:Yuklag}
\end{equation}
The model naturally accommodates a hierarchy of mass scales through its symmetry structure. Although the mass term $M$ is allowed by the $Z_4$ symmetry and can be large, the Majorana mass terms for $N$ and $\chi$, arising from the last two terms in Eq.~\eqref{eq:Yuklag}, are expected to be suppressed either through small Yukawa couplings or a small VEV of $S$. We focus on the former scenario where $\lambda, \lambda' \ll 1$, which is particularly well-motivated from the perspective of symmetry enhancement. The smallness of these couplings is protected by 't Hooft's naturalness principle~\cite{tHooft:1979rat}, as setting $\lambda, \lambda'$ to zero enhances the symmetry of the model to conservation of the lepton number $U(1)_L$, with charge assignments of $+1$ for the leptons ($L$, $e_R$, and $N$) and $-1$ for $\chi$. Quantum corrections to these parameters are proportional to the couplings themselves, preventing large radiative corrections and maintaining their small values. This natural stabilization mechanism makes the scenario technically natural, where the symmetry-enhanced limit protects the hierarchy of scales in this model.

The suppressed mass terms generated through the small couplings $\lambda$ and $\lambda'$ play a dual role. They enable the inverse seesaw mechanism for neutrino mass generation while simultaneously ensuring the feebly interacting nature of the $\chi$ as a DM candidate. The symmetry breaking pattern leads to a mass spectrum where the large mass $M$ dominates over the $\langle S\rangle$-induced small Majorana mass terms, creating the necessary conditions for both neutrino mass generation and freeze-in DM production. The residual $Z_2$ symmetry from the breaking of $Z_4$ ensures the stability of the lightest odd-parity particle, providing a natural candidate for DM. Its feeble interactions with the SM sector are protected by the same symmetry principles that explain the smallness of neutrino masses. This interconnected structure simultaneously addresses two of the most pressing beyond-SM phenomena while maintaining the testability of the model at experimentally accessible energy scales.

\begin{figure}[htbp]
\centering 
\includegraphics[width=0.7\textwidth]{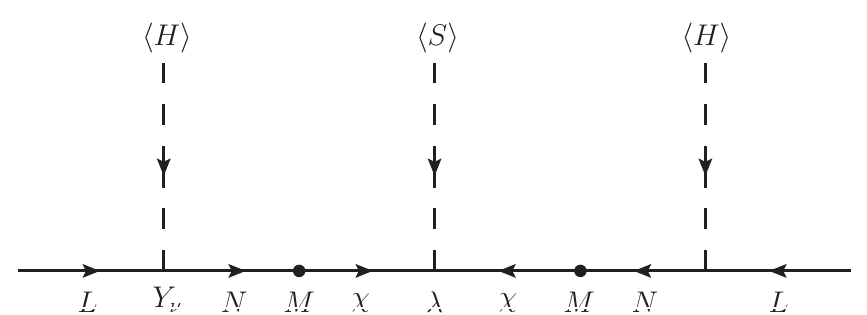} 
\caption{Feynman diagram for light neutrino mass generation via the inverse seesaw mechanism.}
\label{fig:inverse_seesaw} 
\end{figure}

The inverse seesaw mechanism, illustrated in Figure~\ref{fig:inverse_seesaw}, generates neutrino masses through an effective dimension-five Weinberg operator $(LH)(LH)/\Lambda$~\cite{Weinberg:1979sa} after integrating out the heavy fermionic states $N$ and $\chi$. Here, $LH \equiv L_r \varepsilon_{rs} H_s$ ($r,s=1,2$) represents the $SU(2)_L$-invariant contraction between the lepton doublet and the Higgs field. The resulting neutrino mass matrix takes the form
\begin{equation}
    m_\nu \simeq \left( \langle H \rangle^2 \langle S \rangle\right) Y_\nu (M^T)^{-1} \lambda M^{-1} Y^T_\nu \; ,
    \label{eq:nufromEFT}
\end{equation}
where $\langle H \rangle$ and $\langle S \rangle$ denote the VEVs of the Higgs and additional scalar field, respectively. This expression reveals the characteristic double suppression of the inverse seesaw mechanism. That is, the neutrino mass is simultaneously suppressed by the large mass scale $M$ and the small Yukawa coupling $\lambda$, which naturally explains the observed sub-eV neutrino masses.
The advantage of this setup lies in its dual suppression structure. Unlike the conventional seesaw, which demands extremely high scales ($\gtrsim 10^{14}$ GeV) to explain neutrino masses, the inverse seesaw achieves the same result through the combination of a TeV-scale mass $M$ and a small mass term $\mu = \lambda\langle S\rangle$ for $\chi$ fermions. This allows the Dirac Yukawa couplings $Y_\nu$ to remain of $\mathcal{O}(1)$, maintaining naturalness while keeping the new physics within experimental reach. The inverse seesaw thus provides an elegant solution to the neutrino mass problem that is both theoretically compelling and phenomenologically testable.

The scalar potential, invariant under the considered symmetries, is given by
\begin{equation}
\begin{aligned}
V(H,S)=& -\mu^{2}_{H}|H|^{2}-\mu_{S}^{2}|S|^{2}+\frac{\lambda_{H}}{2}|H|^{4}+\frac{\lambda_{S}}{2}|S|^{4}+\lambda_{H,S}|H|^{2}|S|^{2} \\
& +\left(\frac{1}{2}\tilde{\mu}^2 S^2+\frac{\eta_1}{2}|H|^{2}S^2+\frac{\eta_2}{2}|S|^{2}S^2+\frac{\kappa}{4}S^4+\mathrm{h.c.}\right),
\end{aligned}
\end{equation}
where all parameters are taken to be real to preserve CP symmetry. For the potential to be bounded from below, the quartic couplings must satisfy the conditions $\lambda_{H}\ge 0$, $\lambda_{S}+\kappa+2\left|\eta_2\right|\ge 0$, $\lambda_{S} - \kappa \ge \eta^2_2/(2\kappa)$ and $\left| \kappa \right|\ge \left| \eta_2 \right|/2$. The scalar VEVs break different symmetries at distinct energy scales. At higher energies, the singlet scalar acquires a VEV $\langle S \rangle$, spontaneously breaking the $Z_4$ symmetry down to its $Z_2$ subgroup.  Subsequently, at the electroweak scale, the Higgs doublet obtains a VEV $ \langle H \rangle $, breaking the electroweak symmetry. To analyze the symmetry breaking, we expand the Higgs doublet and the singlet scalar around their respective VEVs in the unitary gauge, 
\begin{equation}
H=\frac{1}{\sqrt{2}}
\begin{pmatrix}
0\\
v+\phi_H
\end{pmatrix}, \qquad 
S=\frac{1}{\sqrt{2}}(u+\phi_S) e^{i \varphi},
\end{equation}
where $\langle S \rangle= u/\sqrt{2}$ and $\langle H \rangle = v/ \sqrt{2}$ are the VEVs of $S$ and $H$. The minimization conditions of the potential yield
\begin{align}
\mu_{H}^{2}&=\frac{1}{2} \left[\lambda_{H}v^{2}+ \left(\lambda_{H, S}+\eta_1\right) u^{2} \, \right],\\  
\mu_{S}^{2}&= \frac{1}{2} \left[\left(\lambda_{S}+\kappa + 2\eta_2\right)u^2+\left(\lambda_{H, S}+\eta_1\right) v^{2} +2\tilde{\mu}^2\right].
\end{align}
Deriving these conditions requires evaluating the value $\varphi=\varphi_0$ that minimizes the potential. The minimization with respect to $\varphi$ leads to several possibilities, with $\varphi_0=n\pi/2$ or $\cos2 \varphi_0 = -\frac{1}{2\kappa}\left( \eta_2+ (v^2/u^2)\eta_1 +2\tilde{\mu}^2/u^2\right)$. The global minimum of the potential resides at $\varphi_0 = 0$, provided $2\tilde{\mu}^2+\eta_1 v^2+\eta_2 u^2\le 0$. Other nonzero choices would lead to spontaneous CP violation. The mass-squared matrix for the CP-even scalar fields $(\phi_H, \phi_S)$ is
\begin{equation}
\mathcal{M}^2=
\begin{pmatrix}
\lambda_{H}v^{2} &  \left(\lambda_{H,S}+\eta_1\right)vu\\
 \left(\lambda_{H,S}+\eta_1\right)vu & \left(\lambda_{S}+\kappa+2\eta_2\right)u^{2}
\end{pmatrix}.
\end{equation}
Diagonalizing this matrix gives the mass eigenvalues for the two physical scalars,
\begin{equation}
    m^2_h = \frac{1}{2}\left[\lambda_{H}v^{2}+ \left(\lambda_{S}+\kappa+2\eta_2\right)u^{2}-\sqrt{\left[ \lambda_{H}v^{2}-\left(\lambda_{S}+\kappa+2\eta_2\right)u^{2} \right]^2+4\left(\lambda_{H,S}+\eta_1\right)^2v^2u^2}\; \right],
\end{equation}
\begin{equation}
    m^2_\phi = \frac{1}{2}\left[\lambda_{H}v^{2}+ \left(\lambda_{S}+\kappa+2\eta_2\right)u^{2}+\sqrt{\left[ \lambda_{H}v^{2}-\left(\lambda_{S}+\kappa+2\eta_2\right)u^{2} \right]^2+4\left(\lambda_{H,S}+\eta_1\right)^2v^2u^2}\; \right],
\end{equation}
which correspond to the mass eigenstates $h$ and $\phi$, respectively. Here, $h$ is identified as the SM-like Higgs boson and $\phi$ is a heavier scalar. In the limit $u \gg v$, the masses are approximately
\begin{equation}
   m^2_h \approx \left(\lambda_{H}- \frac{\left(\lambda_{H,S}+\eta_1\right)^2 }{\lambda_{S}+\kappa+2\eta_2} \right) v^2, \qquad    m^2_\phi \approx \left(\lambda_{S}+\kappa+2\eta_2\right)u^{2}+\frac{\left(\lambda_{H,S}+\eta_1\right)^2}{\lambda_{S}+\kappa+2\eta_2}v^2. 
\end{equation}
These expressions show that the standard SM Higgs mass is recovered when the portal couplings $\lambda_{H,S}$ and $\eta_1$ vanish.  The linear transformations between the original scalar fields and the mass eigenstates are given by
\begin{align}
		  h & =\phi_H \cos{\theta}-\phi_S\sin{\theta},
		\\
        \phi &=\phi_S\cos{\theta}+\phi_H \sin{\theta},
\end{align}
where the mixing angle $\theta$  satisfies
\begin{equation}
    \tan \theta = \frac{\left(\lambda_{S}+\kappa+2\eta_2\right)u^{2}-\lambda_{H}v^{2} }{2\left(\lambda_{H,S}+\eta_1\right)vu} \left[\sqrt{1+\frac{4\left(\lambda_{H,S}+\eta_1\right)^2v^2u^2}{\left(\left(\lambda_{S}+\kappa+2\eta_2\right)u^{2}-\lambda_{H}v^{2}\right)^2} } -1\right].
\end{equation}
For $u \gg v$, the mixing angle is small and can be approximated as
\begin{equation}
    \tan \theta \approx \frac{\left(\lambda_{H,S}+\eta_1\right)v }{ \left(\lambda_{S}+\kappa+2\eta_2\right)u} \left[ 1+\frac{v^2}{\left(\lambda_{S}+\kappa+2\eta_2\right)u^2 } \left( \lambda_{H}-  \frac{\left(\lambda_{H,S}+\eta_1\right)^2}{\lambda_{S}+\kappa+2\eta_2} \right)  \right].
\end{equation}
It is clear that the mixing angle $ \theta $ vanishes when the couplings $\lambda_{H,S} $ and $\eta_1$ are set to zero. In this limit, the SM Higgs mass reverts to its standard form, and the CP-even scalar sector consists solely of an additional self-interacting singlet scalar that does not interact with the Higgs boson. Given current experimental constraints, an upper limit on the mixing angle is $\sin \theta < 0.1 $~\cite{Robens:2015gla,Dupuis:2016fda,Bechtle:2020uwn,Lane:2024vur}. Such a small mixing is insufficient to produce observable effects in collider experiments, including Higgs factories.

After spontaneous symmetry breaking, the phase $\varphi$ becomes a pseudo-Nambu-Goldstone boson (pNGB) with mass
\begin{equation}
    m_\varphi = \sqrt{ 2\tilde{\mu}^2 +\left(\eta_2+2\kappa \right)u^2 +\eta_1 v^2}\; u.
\end{equation}
As expected, this mass depends only on the $Z_4$ symmetry breaking parameters. In general, the CP-odd scalar $\varphi$ can mix with the CP-even scalars due to the terms $\left| H\right|^2 S^2$, $\left|S\right|^2 S^2$, $S^4$ and their Hermitian conjugates. However, with the choice $\varphi_0=n\pi/2$, these mixing contributions cancel in the scalar mass matrix, and $\varphi$ remains as a mass eigenstate. Nevertheless, it interacts with the other scalars and with new singlet fermions introduced in the model. This leads to rich collider phenomenology, including pNGB-mediated scattering processes and exotic Higgs decays if $m_\varphi < m_h/2$.

\section{The phenomenological aspects}
\label{sec:pheno}
\subsection{Analysis of neutrino masses and mixing}
\label{subsec:neumasmix}

The Yukawa interactions in Eq.~\eqref{eq:Yuklag} generate neutrino masses through spontaneous symmetry breaking. The resulting $9\times9$ mass matrix takes the block form
\begin{equation}
    \mathcal{L}_{m}=
  \begin{pmatrix}
    \overline{L} & \overline{N^c} & \overline{\chi^c}
  \end{pmatrix}
  \begin{pmatrix}
    0 & m_{D} & 0\\
    m^{T}_{D} & \mu^\prime & M\\
    0 & M^{T} & \mu
  \end{pmatrix}
  \begin{pmatrix}
  	L \\ N \\ \chi
  \end{pmatrix},
\end{equation}
where the entries $m_D = vY_\nu/\sqrt{2}$, $\mu = \lambda u/\sqrt{2}$, and $\mu^\prime = \lambda' u/\sqrt{2}$ are $3\times3$ matrices. The $Z_4$ symmetry naturally enforces the hierarchy $\mu, \mu^\prime \ll m_D \ll M$, which is characteristic of the inverse seesaw mechanism. The mass matrix can be compactly written in terms of four blocks,
\begin{equation}
	\mathcal{M}=\begin{pmatrix}
    0 & m_{D} & 0\\
    m^{T}_{D} & \mu^\prime & M\\
    0 & M^{T} & \mu
 \end{pmatrix}=\begin{pmatrix}
		0 & \mathcal{M}_{D}\\
		\mathcal{M}_{D}^{T} & \mathcal{M}_{N}
	\end{pmatrix},
 \label{eq:masmtrx}
\end{equation}
where $\mathcal{M}_{D} =\begin{pmatrix} m_{D} & 0 \end{pmatrix}$, and 
\begin{equation}
	\mathcal{M}_{N}=\begin{pmatrix}
	\mu^\prime & M\\
	M^{T} & \mu
\end{pmatrix}.
\label{eq:heavysectmasmtrx}
\end{equation}
To determine the light neutrino mass matrix within the seesaw framework, it is necessary to compute the inverse of the $\mathcal{M}_{N}$ block. Assuming that $M$ is non-singular, its inverse is
\begin{equation}
    \mathcal{M}_{N}^{-1} =
    \begin{pmatrix}
    -\left(M^T-\mu M^{-1} \mu^\prime  \right)^{-1} \mu M^{-1} & \left(M^T-\mu M^{-1} \mu^\prime  \right)^{-1} \\
     \left(M-\mu^\prime (M^T)^{-1} \mu  \right)^{-1}  & -M^{-1} \mu^\prime \left(M^T-\mu M^{-1} \mu^\prime  \right)^{-1}
  \end{pmatrix}.
\end{equation}
Applying the block diagonalization procedure, the light neutrino mass matrix is obtained as 
\begin{equation}
   m_\nu = -\mathcal{M}_{D}
      \mathcal{M}_{N}^{-1}\mathcal{M}_{D}^{T}=  m_{D} \left(M^T-\mu M^{-1} \mu^\prime  \right)^{-1} \mu \, M^{-1} m^T_{D}
\end{equation}
Given the hierarchy $\mu, \mu^\prime \ll M$, we retain only the leading order terms in $\mu$ and $ \mu^\prime$, neglecting the second term within the parentheses. This simplification yields
\begin{equation}
    m_\nu =  m_{D} \left(M^T\right)^{-1} \mu \, M^{-1} m^T_{D},
    \label{eq:lghtnumas}
\end{equation}
which is consistent with the result obtained from the effective field theory approach in Eq.~\eqref{eq:nufromEFT}. The neutrino mass matrix $m_\nu$ is proportional to the small mass parameter $\mu$ and is further suppressed by the heavy mass scale $M$, characteristic of the inverse seesaw mechanism. It should be emphasized that this result can be easily extended to scenarios involving more than three generations. In contrast, it is equally applicable to cases with only one or two generations. For the single-generation case, the mass matrix simplifies to
\begin{equation}
    m_\nu =  \frac{y_\nu^2}{2M^2}  v^2 \mu,
\end{equation}
which yields sub-eV neutrino masses for a Majorana mass term $\mu\sim \mathcal{O}(1)$ MeV and $M\sim \mathcal{O}(10)$ TeV with $\mathcal{O}(1)$ Yukawa coupling.
 
The block diagonalization of the mass matrix  $\mathcal{M}$ in Eq.~\eqref{eq:masmtrx} is performed using the rotation matrix
\begin{equation}
    \mathcal{U} =
    \begin{pmatrix}
        \mathbf{I}_{3} - \frac{1}{2} \epsilon \epsilon^T & \epsilon \\
        -\epsilon^T & \mathbf{I}_{3} - \frac{1}{2} \epsilon^T \epsilon
    \end{pmatrix},
    \label{eq:blckunitry}
\end{equation}
where $\epsilon =\mathcal{M}_D \mathcal{M}^{-1}_N= (-m_\nu (m^T_D)^{-1}, m_D(M^T)^{-1})$. By this rotation, the mass matrix gets block diagonalized as
\begin{equation}
   \mathcal{U}^T  \mathcal{M} \; \mathcal{U} = \begin{pmatrix}
       -\mathcal{M}_D \mathcal{M}^{-1}_N \mathcal{M}^T_D  &   \\
             &  \mathcal{M}_N + \frac{1}{2} \left[\mathcal{M}_D^T \left(  \mathcal{M}_D \mathcal{M}^{-1}_N \right)+\left(  \mathcal{M}_D \mathcal{M}^{-1}_N \right)^T \mathcal{M}_D \right]
   \end{pmatrix},
   \label{eq:blkdgnlztn}
\end{equation}
in the leading order of $\mathcal{M}_D \mathcal{M}^{-1}_N$. The mixing between the light and heavy sectors is characterized by the small parameter $\epsilon$, and each diagonal block in Eq.~\eqref{eq:blkdgnlztn} isolates the light and heavy mass spectra. The small corrections to $\mathcal{M}_N$ can be safely neglected in subsequent analyses.  

To fully diagonalize the mass matrix, further unitary transformations are performed on each block. Namely, two additional unitary matrices $ V_\nu $ and $V_N$ are introduced such that
\begin{equation}
    \begin{aligned}
    V_\nu^T m_\nu V_\nu &= \text{diag}(m_1, m_2, m_3), \\
    V_N^T \mathcal{M}_N V_N &= \text{diag}(M_1, \dotsc, M_6),
\end{aligned}
\end{equation}
where $m_1,m_2,m_3$ are the masses of the light neutrinos and $M_1,\dotsc,M_6$  are the masses of the heavy singlet states. 

The neutrino mixing matrix $U_{\rm PMNS}$ is then given by
\begin{equation}
U_{\rm PMNS} = U^\dagger_l (\mathbf{I}_{3} - \frac{1}{2}\epsilon\, \epsilon^T) V_\nu,
\label{eq:PMNSmatrx}
\end{equation}
where $U_l$ diagonalizes the Hermitian combination of the charged lepton mass matrix, satisfying $U^\dagger_l m_E m_E^\dagger U_l = \diag (m^2_e, m^2_\mu, m^2_\tau)$. The term $\frac{1}{2}\epsilon\, \epsilon^T$ quantifies the deviation from unitarity of the mixing matrix. However, since this deviation is quadratic in $\epsilon$, its contribution is negligible and can be ignored in the following discussion.

The heavy singlet fermion masses are determined by diagonalizing the mass matrix $\mathcal{M}_N$ through a unitary transformation $Q^T \mathcal{M}_N Q = D$, where $D$ contains the physical mass eigenvalues. Given the natural hierarchy $\mu, \mu^\prime \ll M$, we apply a perturbative diagonalization approach (detailed in Appendix~\ref{app:srpertblock}) that reveals several characteristic features of the spectrum. At leading order, the diagonalization shows significant mixing between the $N$ and $\chi$ states and three nearly degenerate pairs of mass eigenstates. This degeneracy persists when first-order perturbations in $\mu M^{-1}$ and $\mu^\prime M^{-1}$ are included, implying the robustness of the pseudo-Dirac structure. The degeneracy is lifted only at the second order in perturbation theory, producing small mass splittings $\Delta M_i \sim \mathcal{O}(\mu^2/M)$ between the components of each pair. This spectrum of pseudo-Dirac heavy neutrinos with small mass splittings represents one of the distinctive signatures of the inverse seesaw framework.

The determination of neutrino masses and mixing parameters requires the diagonalization of the mass matrix given in Eq.~\eqref{eq:lghtnumas}. To properly account for the leptonic mixing, we must compute the rotation matrices $U_l$ and $V_\nu$ that diagonalize the charged lepton and neutrino mass matrices respectively. To achieve this in a theoretically constrained framework, we construct a flavor symmetry model based on the simplest non-Abelian discrete group, $S_3$. This symmetry group is defined by two generators $S$ and $T$ satisfying the presentation rules $S^2 = T^2 = (ST)^3 = e$, where $e$ denotes the identity element. This group has two one-dimensional (denoted as $1$ and $1'$) and one two-dimensional irreducible representations $2$. The tensor products between them can be decomposed into irreducible components,
\begin{equation}
    \begin{aligned}
        & 1\times r = r, \quad r\in \{1,~1',~2\}, \\
        & 1' \times 1' =1, \quad 1' \times 2 =2 , \\
        & 2 \times 2 =1 + 1'+2.
    \end{aligned}   
\end{equation}
In particular, in the $\rho(T)$-diagonal basis, the Clebsch-Gordan coefficients for products of two doublets $(x_1, y_1)$ and $(x_2, y_2)$ are given by~\cite{Ishimori:2010au,Novichkov:2019sqv}
\begin{equation}
    \begin{aligned}
       1 \sim x_1 y_1 + x_2 y_2, \quad 
       1' \sim x_1 y_2- x_2 y_1, \quad 
       2 \sim \begin{pmatrix}
           x_1 y_2 + x_2 y_1 \\
           x_1 y_1 - x_2 y_2
       \end{pmatrix}.
    \end{aligned}
\end{equation}
The field assignments, $L_1 \sim 1'$, $L_D \equiv \left(L_2,L_3\right)\sim 2$, $ e_{1,R}\sim 1'$, $\ell_{ R}\equiv \left(e_{2,R},e_{3,R}\right)\sim 2$, and a flavon $\varphi = \left(\varphi_1, \varphi_2\right)\sim 2$, constrain the charged lepton Yukawa Lagrangian
\begin{equation}
    \mathcal{L}_E = Y_{E11} (\overline{L}_1 H)  e_{1,R} + Y_{E22}\left[(\overline{L}_D H ) \ell_{R}\right]_1+\frac{Y_{E23}}{\Lambda} \left[(\overline{L}_D H ) \ell_{R} \varphi \right]_1 + \mathrm{h.c.},
\end{equation}
where $\Lambda$ is the cutoff scale of the flavor model, and the subscript $1$ of the square bracket denotes the singlet combination. To further constrain the mass matrices, an additional $Z_2$ symmetry is imposed, under which $L_1$, $e_{1,R}$ and $N_3$ are odd,  while all other fields are even. Given the VEV alignment $\langle \varphi \rangle = v_\varphi \left(1,0  \right)^T$, it leads to the charged lepton mass matrix
\begin{align}
     m_E   = \begin{pmatrix}
        a_1 & 0 & 0 \\
        0  &  a_2 & -a_3 \\
        0 & -a_3 & a_2 
    \end{pmatrix},
    \label{eq:chrgdlptmss}
\end{align}
where $a_1 = Y_{E11} \langle H \rangle$, $a_2 = Y_{E22} \langle H \rangle$ and $a_3= - Y_{E23} \langle H \rangle v_\varphi/\Lambda$. The charged lepton masses are extracted by the diagonalization $U_l^\dagger m_E U_r = \diag(m_e, m_\mu, m_\tau) $, in which
\begin{equation}
\begin{aligned}
     U_l &  = \begin{pmatrix}
        1 & 0 & 0\\
        0 & \frac{1}{\sqrt{2}} & -\frac{1}{\sqrt{2}} \\
        0  & \frac{1}{\sqrt{2}}  & \frac{1}{\sqrt{2}}
    \end{pmatrix}, \quad U_r = U_l P, \\
    m_e & = |a_1|, \quad m_\mu = |a_2-a_3|, \quad m_\tau = |a_2+a_3|,
\end{aligned}
\end{equation}
with $P =\diag (\sign(a_1), \sign(a_2-a_3), \sign(a_2+a_3))$.

For the neutrino sector, we assign the representations $N_D\equiv \left( N_1, N_2 \right)\sim 2$, $N_3, \chi_1 \sim 1'$, $\chi_D \equiv \left( \chi_2, \chi_3 \right) \sim 2$. The resulting Yukawa Lagrangian responsible for neutrino masses takes the form
\begin{equation}
\begin{aligned}
    \mathcal{L}_\nu = &  Y_{\nu 13}\overline{L}_1 N_3\tilde{H}+Y_{\nu 21}\overline{L}_D N_D\tilde{H}+\frac{Y_{\nu 22}}{\Lambda} \left[\overline{L}_D N_D\tilde{H}\varphi\right]_1  +\lambda_{11} S\, \overline{\chi^c_1}\chi_1 +\lambda_{22} S\, \overline{\chi^c_D}\chi_D \\
    &+\frac{\lambda_{13}}{\Lambda} S\, \overline{\chi^c_1} \left[\chi_D \varphi\right]_{1'}  +\frac{\lambda_{13}}{\Lambda} S\,  \left[\overline{\chi^c_D} \varphi\right]_{1'}\chi_1 +\frac{\lambda_{23}}{\Lambda} S\, \left[\overline{\chi^c_D}\chi_D \varphi\right]_1 \\
     & +M_{31} \overline{N^c_3}\chi_1+M_{12} \overline{N^c_D}\chi_D+ \mathrm{h.c.}.
\end{aligned}
\label{eq:s3constranedyuklag}
\end{equation}
The subscript indices for the terms involving $\varphi$ indicate the specific $S_3$ singlet combinations. This Lagrangian results in the following mass matrix structures
\begin{equation}
\begin{aligned}
    m_D= & \frac{v}{\sqrt{2}} \begin{pmatrix}
        0  & 0  &  Y_{\nu 13} \\
        Y_{\nu 21} &  Y_{\nu 22} \frac{v_\varphi}{\Lambda}  &  0  \\
        Y_{\nu 22} \frac{v_\varphi}{\Lambda}  & Y_{\nu 21} & 0 
    \end{pmatrix}, \quad
    M =  \begin{pmatrix}
        0  & M_{12}  &  0 \\
        0 &  0  &  M_{12}  \\
        M_{31}  & 0  & 0 
    \end{pmatrix}, \quad \\   
    \mu= &  \frac{u}{\sqrt{2}} \begin{pmatrix}
        \lambda_{11}  & 0  &  -\lambda_{13}\frac{v_\varphi}{\Lambda} \\
          0  &  \lambda_{22}  &  \lambda_{23}\frac{v_\varphi}{\Lambda}  \\
        -\lambda_{13}\frac{v_\varphi}{\Lambda} & \lambda_{23}\frac{v_\varphi}{\Lambda}  & \lambda_{22} 
    \end{pmatrix}.    
\end{aligned}
\end{equation}
In these expressions, we neglect the flavor-induced corrections to the mass matrix $M$ due to the large magnitude of its entries, corresponding to the heavy masses. As a result, the light neutrino mass matrix is given by
\begin{equation}
\renewcommand{\arraystretch}{1.8}
    m_\nu = \frac{u v^2 }{2\sqrt{2}}\begin{pmatrix} 
        \frac{Y_{\nu 13}^2 \lambda_{11}}{M^2_{31}} & -\frac{Y_{\nu 13}Y_{\nu 22}\lambda_{13}}{M_{12}M_{31}} r^2 & -\frac{Y_{\nu 13}Y_{\nu 21}\lambda_{13}}{M_{12}M_{31}} r \\ 
        -\frac{Y_{\nu 13}Y_{\nu 22}\lambda_{13}}{M_{12}M_{31}} r^2 & \frac{Y_{\nu 21}^2 \lambda_{22} +(Y_{\nu 22}^2 \lambda_{22}+2Y_{\nu 21} Y_{\nu 22} \lambda_{23})r^2 }{M^2_{12}}& \frac{2Y_{\nu 21} Y_{\nu 22}\lambda_{22}+Y_{\nu 21}^2\lambda_{23}+ r^2 Y_{\nu 22}^2 \lambda_{23}  }{M^2_{12}}r \\ 
        -\frac{ Y_{\nu 13}Y_{\nu 21}\lambda_{13}}{M_{12}M_{31}} r & \frac{2Y_{\nu 21} Y_{\nu 22}\lambda_{22}+Y_{\nu 21}^2\lambda_{23}+ r^2 Y_{\nu 22}^2 \lambda_{23}  }{M^2_{12}}r & \frac{Y_{\nu 21}^2 \lambda_{22} +(Y_{\nu 22}^2 \lambda_{22}+ Y_{\nu 21} Y_{\nu 22} \lambda_{23})r^2  }{M^2_{12}}
    \end{pmatrix},
\end{equation}
where $r= \frac{v_\varphi}{\Lambda}$. 

These constraints of Yukawa couplings not only help to reduce the number of free parameters in the model but also lead to specific, testable predictions for the neutrino mixing angles and mass ordering. For illustrative purposes, we perform a numerical analysis using the \texttt{FlavorPy} package~\cite{FlavorPy,Novichkov:2020eep}, and present our results alongside recent experimental observations~\cite{Esteban:2020cvm}. To this end, we assume $10~\text{TeV}\leq M_{12},M_{31}\leq 10^3~\text{TeV}$, $1~\text{TeV}\leq u\leq 10^2~\text{TeV}$ and make a scan of $0.01 \le r \le 10$. The Yukawa couplings vary in $0.01 \le Y_\nu \le 10$, and the coupling $\lambda$ lies in the range $10^{-13} \le \lambda \le 10^{-10}$, keeping consistent within the values used to account for the DM relic density (see discussions in Section~\ref{subsubsec:Dmrd}). Due to the large $\chi^2$ values and consequent poor agreement with experimental data obtained for the case of inverted mass ordering, we restrict our subsequent analysis to the normal ordering case.
\begin{figure}[htbp]
\centering 
\includegraphics[width=0.6\textwidth]{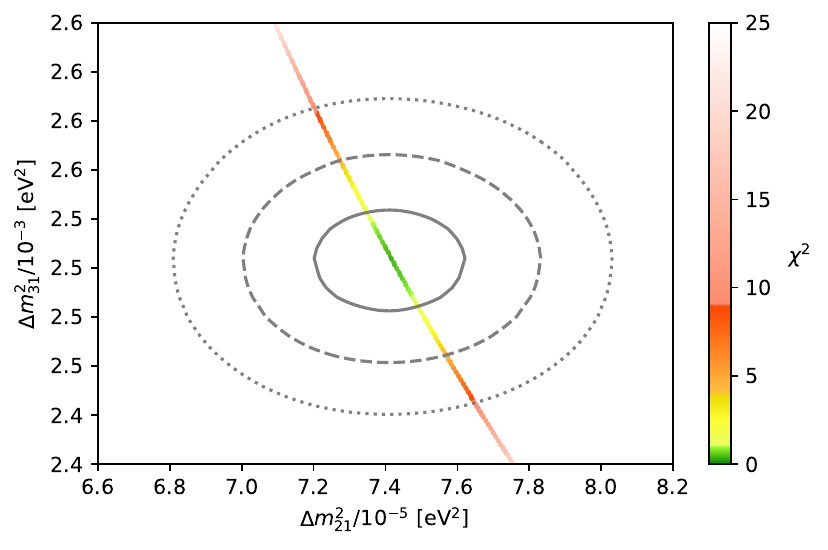}
\caption{Correlation between the neutrino mass-squared differences $\Delta m^2_{31}$ and $\Delta m^2_{21}$. The solid, dashed, and dot-dashed lines in the background represent the $1\sigma$, $2\sigma$, and $3\sigma$ experimental uncertainty regions respectively. The color bar on the right indicates $\chi^2$ values. The cooler colors correspond to better agreement, as they have smaller $\chi^2$ values.}
\label{fig:M2} 
\end{figure}
Figure~\ref{fig:M2} illustrates the correlation between the two neutrino mass-squared differences $\Delta m^2_{31}$ and $\Delta m^2_{21}$, indicating an approximately inverse proportional relationship. Furthermore, the theoretical predictions show good consistency with the experimental values, with many points falling within the $1\sigma$ confidence region.
\begin{figure}[htbp]
\centering 
\includegraphics[width=1.0\textwidth]{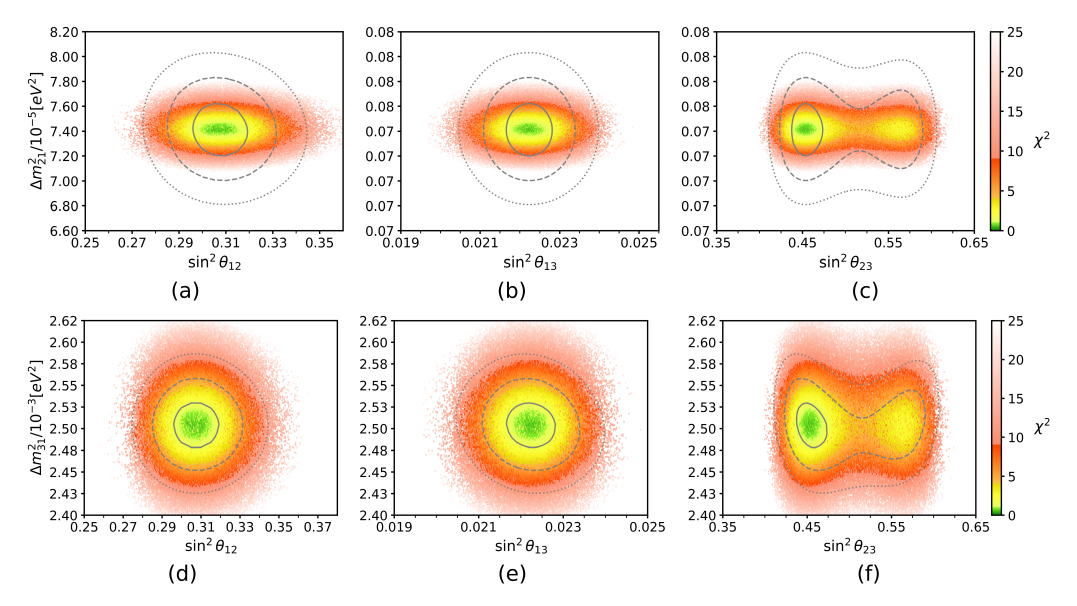}
\caption{Correlations between the neutrino mass-squared differences and the mixing angles. Panels (a), (b), and (c) show $\Delta m^2_{21}$ versus $\sin^2 \theta_{12}$, $\sin^2 \theta_{13}$, and $\sin^2 \theta_{23}$, respectively; panels (d), (e), and (f) show $\Delta m^2_{31}$ versus the same three angles. The gray contours indicate the experimental confidence regions as in Figure~\ref{fig:M2}, with color scale representing $\chi^2$ values.}
\label{fig:M1} 
\end{figure}

Figure~\ref{fig:M1} shows the correlations between the neutrino mass-squared differences and mixing angles. The gray solid, dashed, and dotted lines in the background denote the $1\sigma$, $2\sigma$, and $3\sigma$ experimental uncertainty regions respectively, the same as in Figure~\ref{fig:M2}. 
Subfigures (a)-(c) show the dependence of $\Delta m^2_{21}$ on each sine-square of the three mixing angles. As evident from these plots, there is excellent agreement within the experimentally allowed ranges. Subfigures (d)-(f) display similar correlations for $\Delta m^2_{31}$, covering the full $3 \sigma$ range. 
\begin{figure}[htbp]
\centering 
\includegraphics[width=1.0\textwidth]{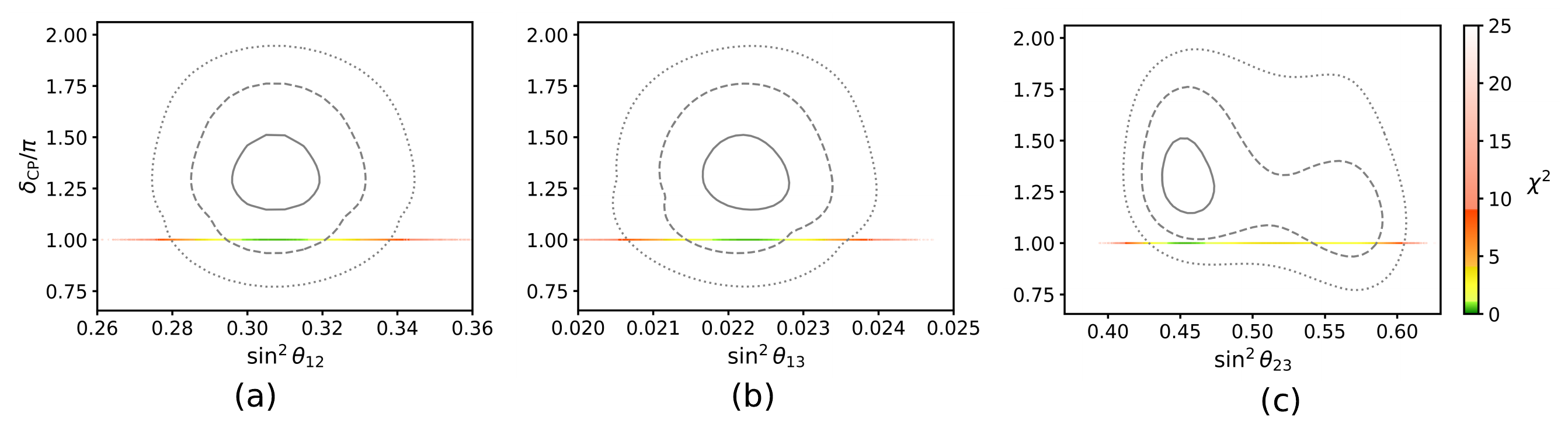}
\caption{Relationships between the Dirac \textit{CP}-violating phase and the leptonic mixing angles: panel (a) is for $\sin^2 \theta_{12}$, (b) is for $\sin^2 \theta_{13}$, and (c) is for $\sin^2 \theta_{23}$. The color mapping is consistent with previous figures.}
\label{fig:cp_theta} 
\end{figure}

The CP-violating phase behavior shown in Figure~\ref{fig:cp_theta} indicates that $\delta_{CP}$ remains relatively stable near $\pi$ across the ranges of all three mixing angles. While the model predicts a broader range for $\sin^2 \theta_{23}$ compared to the other angles, all values remain within the $3\sigma$ confidence level of current experimental results.

\begin{figure}[htbp]
\centering 
\includegraphics[width=1.0\textwidth]{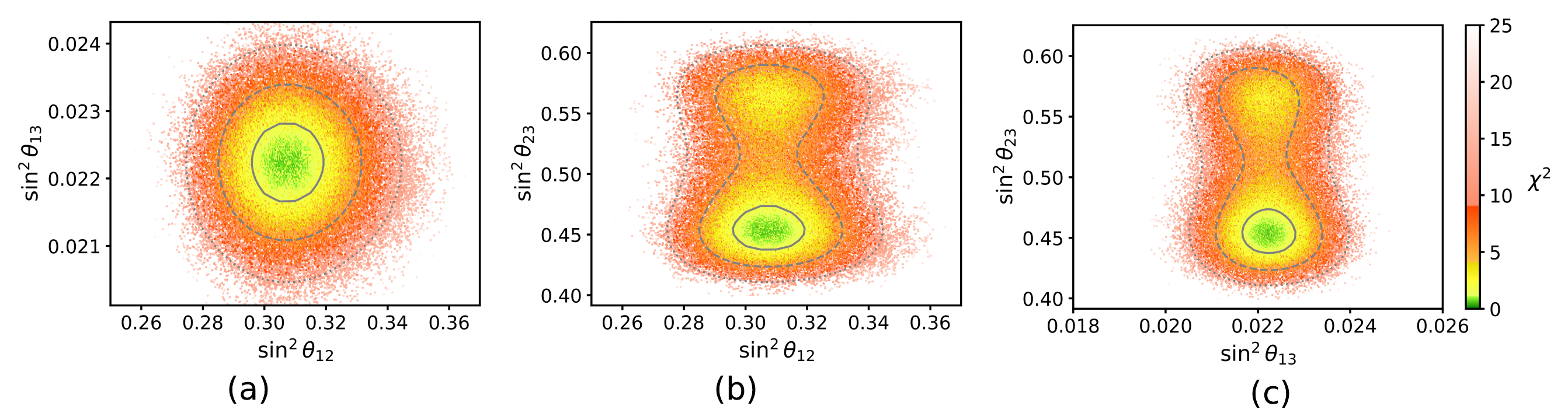}
\caption{The correlations between mixing angles: (a) is for $\sin^2 \theta_{12}$ vs $\sin^2 \theta_{13}$, (b) is for $\sin^2 \theta_{12}$ vs $\sin^2 \theta_{23}$, and (c) is for $\sin^2 \theta_{13}$ vs $\sin^2 \theta_{23}$. The parameter ranges, plotting styles, and the definition of the experimentally allowed regions match previous figures.}
\label{fig:thetas} 
\end{figure}

Figure~\ref{fig:thetas} demonstrates strong agreement across the relationships between the sine-squared values of any two of the three mixing angles. The theoretical values span the full experimental range from $1\sigma$ to $3\sigma$. The consistent agreement across all pairwise angle correlations suggests theoretical consistency in the treatment of mixing parameters.

These comprehensive analyses demonstrate that the framework successfully reproduces the observed neutrino oscillation parameters while making specific predictions for currently less-constrained quantities like $\delta_{CP}$, which remains nearly constant around $\pi$. This is testable in future neutrino oscillation experiments. The remaining deviations in certain parameter spaces, particularly involving $\theta_{23}$, suggest directions for further theoretical refinement, while not invalidating the overall framework. In the following subsection, we investigate additional phenomenological implications, simultaneously accommodating the DM constraints while keeping consistent with neutrino oscillation data. The interplay with the DM sector further strengthens the phenomenological implications of this work for cosmological studies.

\subsection{Freeze-in DM}
\label{subsec:FIMPDM}

The additional singlet fermions $N$ and $\chi$, being electrically neutral and odd under the unbroken $Z_2$ symmetry, possess the necessary properties to serve as viable DM candidates. The lightest $Z_2$-odd fermion remains stable due to the preserved discrete symmetry, making it a compelling relic particle. Crucially, the mixing between these fermions and SM neutrinos, characterized by the small parameter $\epsilon$, is sufficiently suppressed to prevent thermal equilibrium in the early universe. Consequently, rather than following conventional freeze-out dynamics, these particles gradually populate the DM abundance through freeze-in production~\cite{McDonald:2001vt,Hall:2009bx,Bernal:2017kxu}. In this scenario, the heavy singlets $N$ and $\chi$ decay or annihilate into the lightest singlet $\chi$, which serves as a FIMP DM candidate. Our analysis assumes negligible initial abundance for $\chi$, with its relic density being generated entirely through these out-of-equilibrium processes. In the following sections, we provide a detailed analysis of the production mechanisms of the FIMP $\chi$ and compute its relic density.

\subsubsection{Decay-assisted DM Production}

\begin{figure}[htbp]
\centering 
\includegraphics[width=0.75\textwidth]{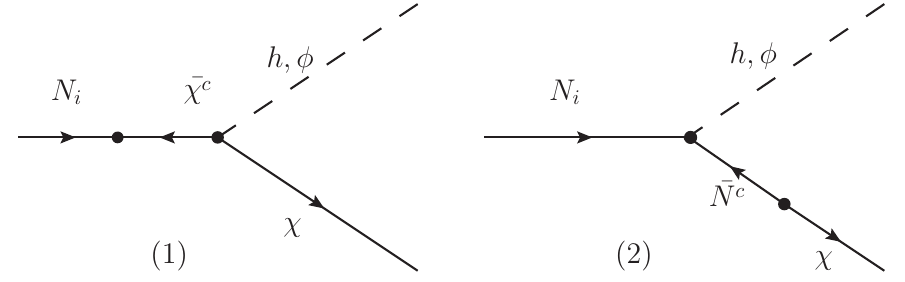} 
\caption{Feynman diagrams for DM production via decays of right-handed neutrinos $N_i$.}
\label{fig:fimpprodfrdcy} 
\end{figure}

Primary production of the DM candidate $\chi$ proceeds via heavy neutrino decays $N_i \to \chi h$ and $N_i \to \chi \phi$, as illustrated in Figure~\ref{fig:fimpprodfrdcy}. These processes become significant due to the Yukawa coupling between $\chi$ and $N$ combined with their substantial mixing. The resultant DM yield is given by~\cite{Hall:2009bx,Belanger:2018ccd}
 \begin{equation}
     Y_{\chi} \simeq 0.656 \frac{M_{\rm Pl} \left( \Gamma_{N_i \to h \chi}  +\Gamma_{N_i \to \phi \chi}\right) }{g_{*S}\sqrt{g_{*}}M_{N_i}^2},
 \end{equation}
where $g_{*}$ and $g_{*S}$ represent the effective numbers of relativistic degrees of freedom for energy density and entropy density respectively, and $M_{\rm Pl} $ is the Planck mass. The sum of decay rates $\Gamma_{N_i \to \chi h}$ and $ \Gamma_{N_i \to \chi \phi}$ is calculated as
 \begin{equation}
     \Gamma_{N_i \to h \chi}  +\Gamma_{N_i \to \phi \chi}=\frac{(m_N^4-m_\chi^4)}{2^8\pi m_N^3}\left[\lambda^2+\lambda^{\prime 2}+\lambda\lambda^\prime(1+\sin 2\theta)\right],
 \end{equation}
where $m_{N}$ and $m_\chi$ are the masses of the heavy neutrino $N$ and the DM candidate $\chi$, while $\lambda$ and $\lambda^{\prime}$ denote the Yukawa couplings. In this analysis, maximal mixing between $N$ and $\chi$ is assumed. This expression implies that 
the kinematic threshold at $m_N = m_\chi$ suppresses these decays. Moreover, the scalar mixing angle $\theta$ contributes minimally due to its small magnitude.
We find that the decay of heavy neutrinos $N_i$ into the DM $\chi$ and scalar particles $h$ or $\phi$ provides a significant production channel. The DM yield $Y_{\chi}$ is computed in detail by incorporating the thermal history of the early universe, in accordance with the freeze-in production paradigm~\cite{McDonald:2001vt,Hall:2009bx,Bernal:2017kxu}.

 \subsubsection{DM production through scattering}

\begin{figure}[htbp]
\centering 
\includegraphics[width=0.8\textwidth]{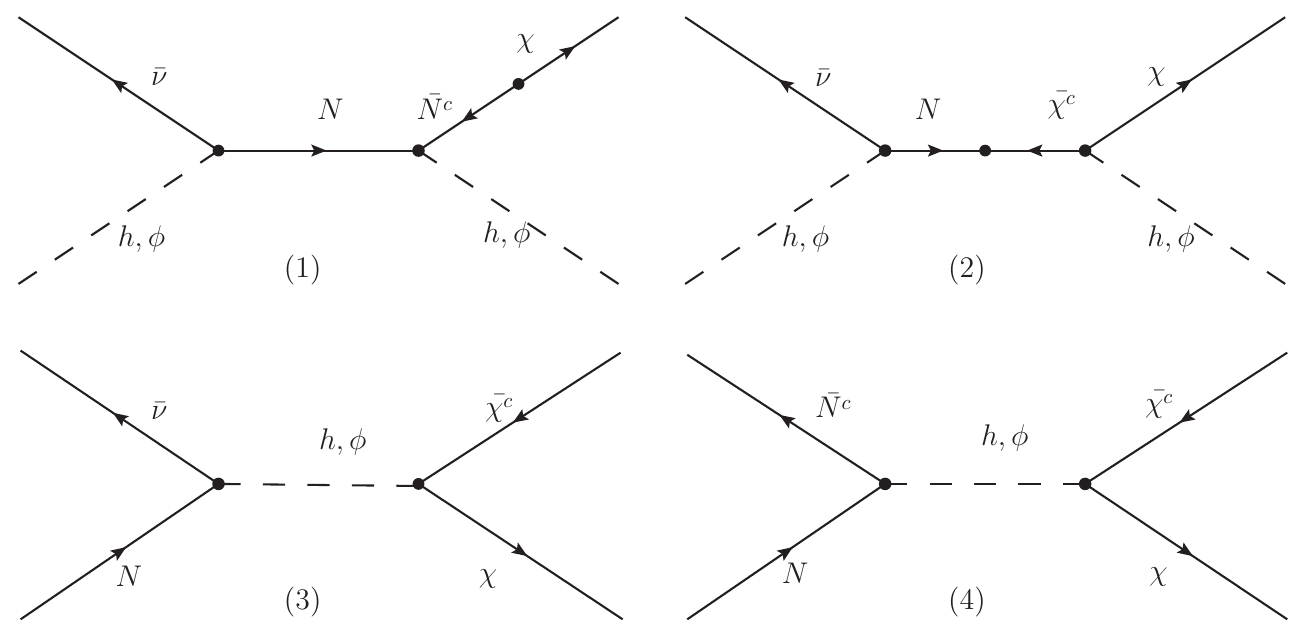} 
\caption{Feynman diagrams for DM production via scattering processes.}
\label{fig:fimpprodfrscttr} 
\end{figure}

Complementary production of $\chi$ occurs through $2\to 2$ scattering processes, as shown in Figure~\ref{fig:fimpprodfrscttr}. These include $\bar{\nu}h~(\text{or}~\phi) \to \chi h~(\text{or}~\phi)$, $\bar{\nu} N \to \chi \bar{\chi}^c$ and $\bar{N}^c N \to \chi \bar{\chi}^c$ scatterings. The first two diagrams involve single DM particle production, while the latter two correspond to pair production.  

As is well-known, the Boltzmann equation governing the time evolution of the DM number density $n_\chi$ gives
\begin{equation}
    \frac{d n_{\chi}}{d t} +3 H n_{\chi} = \int \frac{d^3 p_a}{(2\pi)^3 2 E_a} \frac{d^3 p_b}{(2\pi)^3 2 E_b} \frac{d^3 p_\chi}{(2\pi)^3 2 E_\chi} \frac{d^3 p_c}{(2\pi)^3 2 E_c} \, (2\pi)^4 \delta^4(p_a + p_b - p_\chi - p_c) |\mathcal{M}|^2 f_a f_b,
\end{equation}
where $p_i$ and $E_i$ denote the four-momenta and energies of the participating particles for the scattering process $a + b \rightarrow \chi + c$, and $f_a$ and $f_b$ are the distribution functions of the initial states. The right-hand side represents the collision term, which quantifies the rate at which DM particles are produced in the early universe through scattering processes, assuming that the initial DM abundance is negligible. The collision term can be expressed in terms of the number densities of the initial states and the thermally averaged cross section,
\begin{equation}
    C = \int \frac{d^3 p_a}{(2\pi)^3} \frac{d^3 p_b}{(2\pi)^3} f_a f_b \sigma v_{\text{rel}} \simeq n_a n_b \langle \sigma v \rangle.
\end{equation}

Given the substantial mass of the DM candidate, single DM production is feasible only when the initial state energy exceeds the DM mass. In contrast, pair production requires the initial energy to surpass twice the DM mass. When analyzing the contributions of these processes, it is reasonable to neglect the contributions from scalars and light neutrino masses. This simplification not only renders the computations more tractable, but also provides an effective approximation. In the subsequent calculations, we treat these processes separately.

The scattering cross sections for the processes (1) and (2), mediated by a heavy singlet, are computed, and their contributions are summed
\begin{equation}
     \sigma(1+2)= \frac{(\lambda + \lambda^{\prime })^2}{2^{11}\pi } |Y_\nu|^2 \frac{\cos^2(2\theta) m_N^2}{(s - m_N^2)^2},
\end{equation}
where $s$ is the squared center-of-mass energy. The corresponding thermally averaged cross section  is
\begin{equation}
\begin{split}
     \langle \sigma v \rangle(1+2) &= \frac{(\lambda + \lambda')^2 |Y_\nu|^2 \cos^2 (2\theta) m_N^2}{2^{16} \pi T^5} \int_{m_\chi^2}^\infty \frac{s^{3/2}}{(s - m_N^2)^2} K_1 \left(\frac{\sqrt{s}}{T}\right) \, ds \; ,
\end{split}
\end{equation}
where $K_1$ is the modified Bessel function of the second kind of order one. Consequently, the collision term is given by 
\begin{equation}
C(1+2)=\frac{3\zeta(3)^{2}T}{2^{17}\pi^{5}}(\lambda+\lambda^{\prime})^2\left|Y_{\nu}\right|^2\cos^2(2\theta)m_N^2\int_{m_\chi^2}^\infty \frac{s^{3/2}}{(s - m_N^2)^2} K_1 \left(\frac{\sqrt{s}}{T}\right) \, ds.
\end{equation}

As a common practice, the DM yield $Y_{\chi}=\frac{n_\chi}{S}$ is introduced, by which the Boltzmann equation becomes
\begin{equation}
   Y_{\chi}(1+2)=\int^{\infty}_{0}\frac{C(1+2)}{S\operatorname{H}T} dT,
   \label{eq:dmyield}
\end{equation}
where the entropy density $S$ and the Hubble parameter $H$ during the radiation-dominated era are  written as
\begin{equation}
\begin{split}
    S=&\frac{2\pi^{2}}{45}g_{*s}T^{3},\\
  H=&\frac{1.66\sqrt{g_{*}}T^{2}}{M_{{\mathrm{Pl}}}}.
\end{split}
\end{equation}
Substituting these expressions into Eq.~\eqref{eq:dmyield}, the DM yield produced through (1) and (2) is given by
\begin{equation}
    Y_\chi(1+2)=\frac{135\zeta(3)^2 (\lambda+\lambda^\prime)^2|Y_\nu|^2\cos^2{(2\theta)}m_N^2 M_{\text{Pl}}}{1.66\times 2^{18}\pi^7 g_{*s}\sqrt{g_*}} \int_0^\infty\int_{m_\chi^2}^\infty \frac{s^{3/2}}{T^5(s - m_N^2)^2} K_1 \left(\frac{\sqrt{s}}{T}\right) \, dsdT.
\end{equation}

For process (3), the scattering cross section is calculated,
\begin{equation}
    \sigma(3)= \frac{\lambda^2 |Y_\nu|^2}{2^{13 }\pi} \sin^2(2\theta) \sqrt{1 - \frac{4m_\chi^2}{s}}\left( \frac{1}{s - m_\phi^2} - \frac{1}{s - m_h^2} \right)^2 (s - 2 m_\chi^2).
\end{equation}
In this model, we adopt a well-motivated mass hierarchy $m_N, m_\chi, m_\phi \gg m_h \gg m_\nu $. This hierarchy enables a controlled approximation for calculating the thermally averaged cross section. Besides, we consider a non-relativistic heavy right-handed neutrino $N$ and a relativistic neutrino $\bar{\nu}$ in the initial state. Under this condition and the assumed mass hierarchy, we derive the following approximate expression for the thermally averaged cross section
\begin{equation}
    \langle \sigma v \rangle (3)\approx \frac{15 \lambda^2 |Y_\nu|^2 \sin^2(2\theta) m_\phi^4 (m_N^2 - 2 m_\chi^2)}{2^{16} \sqrt{2\pi m_N} m_N^4 (m_N^2 - m_\phi^2)^2} T^{1/2}.
\end{equation}
The corresponding collision term is
\begin{equation}
    C(3) = \frac{45 \zeta(3) \lambda^2 |Y_\nu|^2 \sin^2(2\theta) m_\phi^4 (m_N^2 - 2 m_\chi^2)}{2^{18} \pi^4 m_N^3 (m_N^2 - m_\phi^2)^2} T^5 e^{-m_N / T}.
\end{equation}
Substituting this expression into the Boltzmann equation and performing the integration of temperature gives
\begin{equation}
    Y_\chi(3) = \frac{45^2 \zeta(3) M_\text{Pl} \lambda^2 |Y_\nu|^2 \sin^2(2\theta) m_\phi^4 (m_N^2 - 2 m_\chi^2)}{1.66 \times 2^{19} \pi^6 m_N^3 (m_N^2 - m_\phi^2)^2 g_{*s} \sqrt{g_*}} \int_0^\infty T^{-1} e^{-m_N / T} dT.
\end{equation}
 
The cross section of the process (4) is computed,
\begin{equation}
\begin{split}
        \sigma(4) &= \frac{\lambda^2 \lambda^{\prime 2}}{2^{12}\pi s} \sqrt{\frac{s-4m_\chi^2}{s-4m_N^2}}\left( \frac{\sin^2 \theta}{s - m_h^2} + \frac{\cos^2 \theta}{s - m_\phi^2} \right)^2 (s - 2m_\chi^2)(s - 2m_N^2).
\end{split}
\end{equation}
 Although the terms in the parentheses add up, this process has a non-trivial contribution only when the temperature of the singlet sector is sufficiently high, enabling efficient collisions of two right-handed neutrinos to produce a pair of DM particles. Given that $m_N $ and $ m_\chi$ are significantly large, such that $s \gg m_h^2, m_\phi^2 $, and during freeze-in $m_N \gg T$, we approximate $s\sim 4m_N^2$. Under this approximation, the thermally averaged cross section is calculated,
\begin{equation}
    \langle \sigma v \rangle(4)=\frac{\lambda^2\lambda^{\prime 2}\sqrt{m_\chi (m_N-m_\chi)}}{2^{13}\sqrt{2}\pi m_N^4}\left[ \left( 2 m_\chi - \frac{3 m_\chi^2}{2m_N}  \right)+3T \left(\frac{1}{2} +\frac{2m_\chi}{m_N} - \frac{3m_\chi^2}{2m_N^{2}}   \right)+  \dfrac{15T^2}{2m_N}\right],
\end{equation}
where the relative velocity $\sqrt{s-4m^2_N}/m_N$  of the initial states is used. In this non-relativistic regime, the collision term is expressed as
\begin{equation}
    C(4)=\frac{\lambda^2\lambda^{\prime 2}\sqrt{m_\chi (m_N-m_\chi)}}{2^{14}\sqrt{2}\pi^4 m_N}e^{-\frac{2m_N}{T}}T^3\left[ \left( 2 m_\chi - \frac{3 m_\chi^2}{2m_N}  \right)+3T \left( \frac{1}{2} +\frac{2m_\chi}{m_N} - \frac{3m_\chi^2}{2m_N^{2}}   \right)+  \dfrac{15T^2}{2m_N}\right].
\end{equation}
With this result, after performing the temperature integration over both sides of the Boltzmann equation yields
\begin{equation}
   Y_\chi(4)= \frac{45M_{{\mathrm{Pl}}}\lambda^2\lambda^{\prime 2}(m_N^2-6m_\chi^2+8m_\chi m_N)\sqrt{m_\chi (m_N-m_\chi)}}{1.66\times 2^{17}\sqrt{2}\pi^6 g_{*s}\sqrt{g_{*}}m_N^4}.
\end{equation}
  
Thus far, we have completed all necessary computations to determine the DM relic abundance. In the following sections, we present the complete expressions and proceed with a numerical analysis based on these results.

\subsubsection{DM relic density}
\label{subsubsec:Dmrd}

The total DM relic density, generated from both decay and scattering processes, is given by
\begin{equation}
    \Omega_\chi h^2=\frac{m_\chi S_0 }{\rho_c/h^2}\left(\sum_{i=1}^{4}Y_\chi(i) +Y_\chi  \right),
\end{equation}
where $Y_\chi(i)$ are the yields from the scattering processes computed previously, and $Y_\chi$ denotes the yield from the decay processes. Using this expression, we compare our theoretical predictions with the experimental measurements. For numerical analysis, we adopt the following parameter values: $\rho_c/h^2 = 1.05\times 10^{-5}$ GeV cm$^{-3}$, $Y_\nu = 0.1$, $\theta = 10^{-5}$, $m_\phi = 5$ TeV. 

\begin{figure}[htbp]
	\centering
    \includegraphics[width=0.49\textwidth]{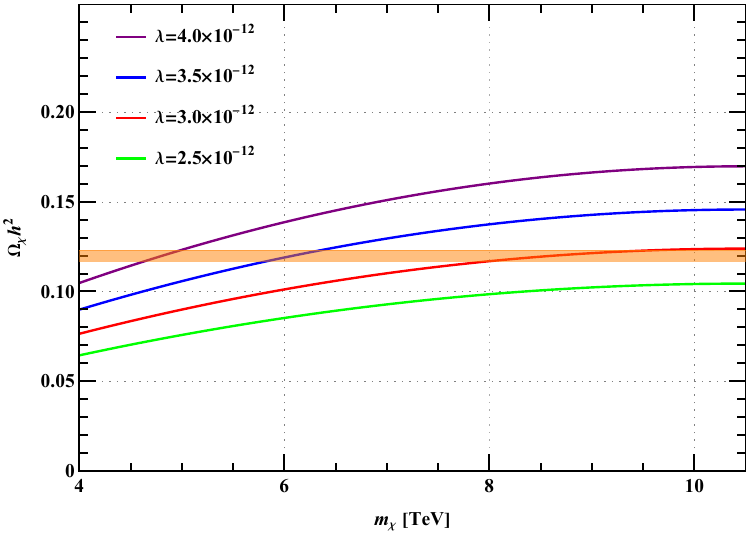}
     \includegraphics[width=0.49\textwidth]{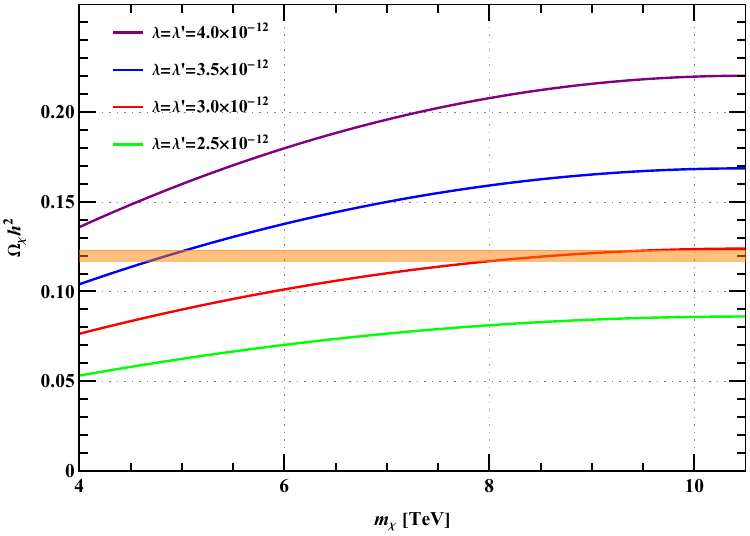}
	\caption{Left: relic density versus DM mass for fixed $\lambda^\prime = 3.0 \times 10^{-12}$ and varying $\lambda$. Right: both couplings are varied equally. The orange band shows the 3$\sigma$ allowed region of the Planck measurement~\cite{Planck:2018vyg}.}
	\label{fig:Omegah2vsparametrs}
\end{figure}

During our numerical study, we observe that the couplings $\lambda$ and $\lambda^\prime$ play the same role to generate the DM relic abundance. That is, exchanging these two couplings leaves the result unchanged. This is also featured in our theoretical expressions. Based on this symmetry, in Figure~\ref{fig:Omegah2vsparametrs} we present the relic density as a function of DM mass for various coupling values, compared with Planck measurements~\cite{Planck:2018vyg}. The left panel illustrates that for fixed $\lambda^\prime = 3.0\times 10^{-12}$, certain mass ranges yield the observed relic density, while the right panel shows equivalent behavior when varying both couplings simultaneously.

In the left panel of Figure~\ref{fig:Omegah2vsparametrs}, we fix $\lambda^\prime = 3.0\times 10^{-12}$ and $m_N =21$ TeV, and plot the DM relic abundance as a function of its mass for different values of the coupling $\lambda$. As illustrated, for a given value of $\lambda$, the DM relic abundance reaches the experimental $3\sigma$ range within certain intervals of the DM mass. In particular, when $\lambda=3.0 \times 10^{-12}$, identical to $\lambda^\prime$, the viable DM mass range is wider, spanning from 8 TeV to 10 TeV. In contrast, for a fixed DM mass, the relic abundance increases with increasing coupling strength, indicating that the abundance of $\chi$ freezes in more effectively with stronger interactions. If $\lambda$ remains constant while $\lambda^\prime$ is varied, the resulting curve for relic abundance remains unchanged due to the symmetric role of these couplings. In the right panel, both $\lambda$ and $\lambda'$ are varied simultaneously with the same value, and the qualitative behavior of the curves mirrors that of the left panel.
\begin{figure}[!ht]
\centering 
\includegraphics[width=0.6\textwidth]{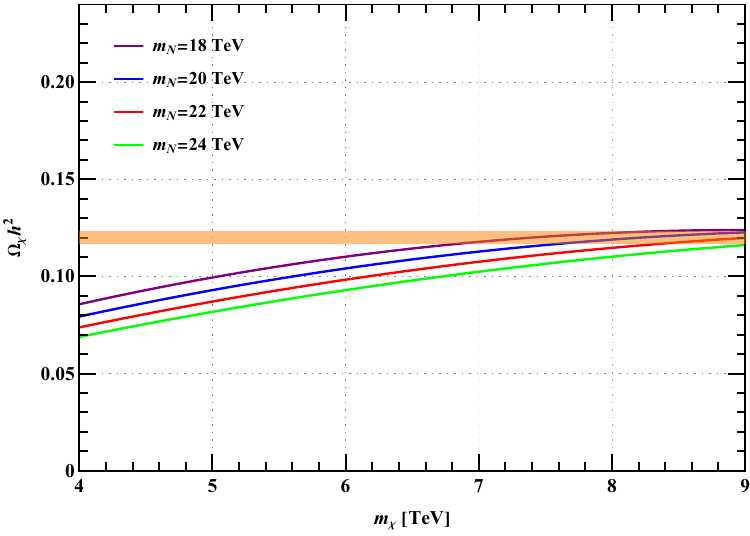}
\caption{DM relic density as a function of its mass for varying $m_N$ values, with fixed couplings $ \lambda = \lambda^\prime = 3.0 \times 10^{-12} $. The $ m_N $ changes from 18 TeV to 24 TeV. Theoretical predictions are compared with the 3$\sigma$ Planck constraint, depicted as the orange band.}
\label{fig:Omegah2_mN} 
\end{figure}

Figure~\ref{fig:Omegah2_mN} shows the DM relic density $\Omega_\chi h^2$ as a function of DM mass for fixed couplings $\lambda=\lambda^\prime = 3.0\times 10^{-12}$, with different curves representing distinct right-handed neutrino masses $m_N$. The plot demonstrates the $m_N$ dependence of the relic density. At a fixed DM mass, $\Omega_\chi h^2$ grows with increasing $m_N$, requiring proportionally larger DM masses to match the observed relic density (orange band). This behavior identifies the viable parameter space for freeze-in production of DM.

Through these analyses, we elucidate the interplay between the DM mass, couplings $\lambda$ and $\lambda^\prime$, and the right-handed neutrino mass $m_N$ to determine the relic density within the freeze-in framework. The figures demonstrate that specific regions of the parameter space yield relic densities consistent with observational data, thereby identifying viable scenarios for DM production in our model.

\section{Summary}
\label{sec:conclusions}
In this paper, we have presented a minimal extension of the SM that simultaneously addresses two fundamental problems in modern physics, the origin of neutrino masses and the nature of DM. The model introduces three right-handed neutrinos, three chiral fermions, and a complex scalar field, all governed by a spontaneously broken $Z_4$ symmetry. This framework tries to demonstrate how discrete symmetries and the inverse seesaw mechanism can provide a unified explanation for both phenomena while maintaining theoretical consistency and phenomenological viability.

Within the inverse seesaw mechanism, the presented model naturally produces small neutrino masses through a double suppression from both the heavy neutrino mass scale ($m_N$) and the small Majorana mass terms ($\mu, \mu'$). The incorporation of $S_3$ flavor symmetry constrains the Yukawa sector, leading to predictive structures in the neutrino mass matrix that successfully reproduce all observed mixing angles within their $3\sigma$ experimental ranges. The model indicates a CP-violating phase $\delta_{\rm CP}\approx\pi$. These correlations and specific predictions will be testable in next-generation neutrino experiments, providing a connection between the high-scale symmetry breaking and the measured neutrino parameters.

The residual $Z_2$ symmetry from $Z_4$ breaking stabilizes  the fermionic DM candidate $\chi$. Its relic abundance is generated through freeze-in mechanisms involving decays ($N_i \to \chi h/\phi$) and scattering processes ($\bar{\nu}N \to \chi\bar{\chi}^c$, etc.). Our systematic phenomenological study identifies viable parameter spaces where the observed DM density emerges naturally from feeble couplings ($\lambda,\lambda' \sim 10^{-12}$) and TeV-scale masses. Furthermore, the interplay between the heavy neutrino masses and the DM mass is mapped out, demonstrating that the model not only aligns with neutrino oscillation parameters but also naturally satisfies cosmological constraints on the DM abundance. 

As a remark, spontaneous breaking of $Z_4$ symmetry inevitably leads to the formation of domain walls, which may give rise to the well-known domain wall problem \cite{Zeldovich:1974uw, Kibble:1976sj, Vilenkin:1981zs, Coulson:1995nv, Krajewski:2021jje}. This issue poses a challenge to the cosmological framework. To circumvent such complications in our model, one may adopt the resolution strategies proposed in Refs.~\cite{Gelmini:1988sf, Kamionkowski:1992mf, Holman:1992us, Larsson:1996sp, Li:2025gld} in the context of $Z_N$ symmetry. Furthermore, other detailed treatments are discussed in Ref.~\cite{Borboruah:2024lli}, where similar methods were used to resolve the domain wall problem arising from spontaneous breaking of the $Z_2$ and $Z_4$ symmetries. These approaches provide a consistent extension to our theoretical framework. This issue is not the central focus of the present work and it is not discussed further.

This work demonstrates that relatively minimal extensions of the SM, guided by the $Z_4$ and $Z_2$ discrete symmetries and the inverse seesaw framework, can successfully unify the explanations for neutrino masses and DM while remaining consistent with all current experimental constraints.  Future work may further investigate the phenomenological implications of the model, including potential signatures in neutrino experiments, collider searches, and DM detection experiments. Moreover, extending the flavor symmetry or introducing additional gauge symmetries may enhance the model's predictive power and improve its consistency with experimental observations. The model's testable predictions and clear theoretical structure make it a rather intriguing scenario for physics beyond the SM.

\section*{Acknowledgements}
This work was supported by the Natural Science Foundation of the Xinjiang Uyghur Autonomous Region of China under Grant No. 2022D01C52. The work of N.Y. was supported by the Natural Science Foundation of the Xinjiang Uyghur Autonomous Region of China under Grant No. 2025D01C292 and by the Doctoral Program of
Tian Chi Foundation of the Xinjiang Uyghur Autonomous Region of China under grant No. 51052300506.


\appendix

\section{Perturbative diagonalization of the heavy sector mass matrix}
\label{app:srpertblock}

We have performed the first block diagonalization in Eqs.~\eqref{eq:blkdgnlztn} and \eqref{eq:blckunitry}, in order to effectively decouple the light and heavy states. Here, we present a detailed perturbative diagonalization of the heavy sector mass matrix $\mathcal{M}_N$ in Eq.~\eqref{eq:heavysectmasmtrx} following the block diagonalization, through which one can determine the singlet fermion masses. Given the mass-scale hierarchy $\mu, \mu^\prime \ll M$, adopting this perturbative diagonalization technique is feasible. To this end, the mass matrix can be decomposed into leading-order and perturbative components
\begin{equation}
     \mathcal{M}_N  \equiv  \mathcal{M}^{(0)}_N  + \mathcal{M}^{(1)}_N   =\begin{pmatrix} 0 & M \\ M^T & 0 \end{pmatrix}+ \begin{pmatrix} \mu^\prime & 0 \\ 0 & \mu \end{pmatrix}.
\end{equation}

First, we diagonalize the leading-order term, and assume that $M$ is a real matrix for simplicity. The latter can be diagonalized by two orthogonal matrices $U_0$ and $V_0$ through singular value decomposition,
\begin{equation}
    M = U_0 D_0 V_0^T,
    \label{eq:singval}
\end{equation}
where $D_0 = \diag(\tilde{M}_1, \tilde{M}_2, \tilde{M}_3)$ is a diagonal matrix with non-negative entries, $\tilde{M}_i \ge 0$. Partitioning the matrices $U_0$ and $V_0$ into column vectors $\vec{u} \in U_0$ and $\vec{v} \in V_0$ , Eq.~\eqref{eq:singval} gives
 \begin{align}
        M \vec{v} = \tilde{M} \vec{u}, \label{eq:srlsvec}\\
        M^T  \vec{u} = \tilde{M} \vec{v}, \label{eq:srrsvec}
    \end{align}
where $\vec{v}$ and $\vec{u}$ are the left and right singular vectors of $M$, respectively. 

Next, we address the eigenvalue problem for $\mathcal{M}^{(0)}_N$. Assuming that its each eigenvector is partitioned into two $3\times 1$ component vectors, $x_i$  and $y_i$, the eigenvalue equation reads
\begin{equation}
    \mathcal{M}^{(0)}_N \begin{pmatrix} \vec{x}_i \\ \vec{y}_i \end{pmatrix} = \lambda_i \begin{pmatrix} \vec{x}_i \\ \vec{y}_i \end{pmatrix},
\end{equation}
which expands to
\begin{equation}
    \begin{pmatrix} 0 & M \\ M^T & 0 \end{pmatrix} \begin{pmatrix} \vec{x}_i \\ \vec{y}_i \end{pmatrix} = \lambda_i \begin{pmatrix} \vec{x}_i \\ \vec{y}_i \end{pmatrix}.
\end{equation}
This is equivalent to two coupled equations
\begin{align}
M \vec{y} = \lambda \vec{x},  \label{eq:etleigvec} \\
M^T \vec{x} = \lambda \vec{y}.  \label{eq:treigvec}
    \end{align}    
By multiplying $M^T$ from the right-hand side of Eq.~\eqref{eq:etleigvec}, we obtain
\begin{equation}
    M^T M \vec{y} = \lambda M^T  \vec{x} = \lambda^2 \vec{y},
\end{equation}
which indicates that $\lambda^2$ are the eigenvalues of $M^T M$. Similarly, from the singular value decomposition, we know that 
\begin{equation}
    M^T M \vec{v}_i = \tilde{M}_i^2 \vec{v}_i.
\end{equation}
This implies that the eigenvalues $\lambda_i$ of $\mathcal{M}^{(0)}_N$ are given by $\lambda_i = \pm \tilde{M}_i$.

Assuming all singular values $\tilde{M}_i$ are nonzero, we first consider the case $\lambda_i = \tilde{M}_i$. The corresponding left eigenvector is $\vec{y}_{(i)} = \vec{v}_{(i)}$, and the associated $\vec{x}_{(i)} $ is
\begin{equation}
\begin{split}
    \vec{x}_{(i)}  &= \frac{1}{\lambda_i} M \vec{y}_{(i)}= \frac{1}{\tilde{M}_i} M \vec{v}_{(i)} = \vec{u}_{(i)}.\\
\end{split}
\end{equation}
Likewise, for $\lambda_i = -\tilde{M}_i$, we find
\begin{equation}
   \vec{x}_{(i)} = \frac{1}{-\tilde{M}_i} M \vec{v}_{(i)} = -\vec{u}_{(i)}.
\end{equation}
Thus, the eigenvectors of $\mathcal{M}^{(0)}_N$ are
 \begin{equation} 
 \begin{cases} 
 \begin{pmatrix} \vec{x}_{(i)} \\ \vec{y}_{(i)} \end{pmatrix} = \begin{pmatrix} \vec{u}_{(i)} \\ \vec{v}_{(i)} \end{pmatrix}, & \text{for} ~\lambda = \tilde{M}_i; \\ 
 \begin{pmatrix} \vec{x}_{(i)} \\ \vec{y}_{(i)} \end{pmatrix} = \begin{pmatrix} -\vec{u}_{(i)} \\ \vec{v}_{(i)} \end{pmatrix}, & \text{for } ~\lambda = -\tilde{M}_i. 
 \end{cases} 
 \end{equation} 
Although the vectors $\vec{u}_{(i)}$ and $\vec{v}_{(i)}$ are individually orthonormal, the combined vectors do not preserve unit normalization,
 \begin{equation}
     \left| \begin{pmatrix} \vec{x}_{(i)} \\ \vec{y}_{(i)} \end{pmatrix} \right|^2 = \left| \vec{x}_{(i)}\right|^2 + \left| \vec{y}_{(i)} \right|^2 = \left| \vec{u}_{(i)} \right|^2 + \left| \vec{v}_{(i)} \right|^2 = 2.
 \end{equation}
To rectify this, we normalize the eigenvectors,
 \begin{equation} 
 \begin{cases} 
 \begin{pmatrix} \vec{x}_{(i)} \\ \vec{y}_{(i)} \end{pmatrix} = \frac{1}{\sqrt{2}} \begin{pmatrix} \vec{u}_{(i)} \\ \vec{v}_{(i)} \end{pmatrix}, & \text{for} ~\lambda = \tilde{M}_i; \\ 
 \begin{pmatrix} \vec{x}_{(i)} \\ \vec{y}_{(i)} \end{pmatrix} = \frac{1}{\sqrt{2}} \begin{pmatrix} -\vec{u}_{(i)} \\ \vec{v}_{(i)} \end{pmatrix}, & \text{for} ~\lambda = -\tilde{M}_i. 
 \end{cases} 
 \end{equation} 
 The matrix $W$ that diagonalizes $\mathcal{M}^{(0)}_N$ is thus constructed
  \begin{equation}
     W = \frac{1}{\sqrt{2}} \begin{pmatrix} U_0 & iU_0 \\ V_0 & -i V_0 \end{pmatrix},
 \end{equation}
 where $U_0$ and $V_0$ are orthogonal matrices composed of the eigenvectors. It is straightforward to verify that $W$ is a $6 \times 6$ unitary matrix, since its columns are orthonormal eigenvectors. This diagonalization procedure yields
 \begin{equation}
      W^T \mathcal{M}^{(0)}_N W = D_0 \oplus  D_0.
 \end{equation}
This result indicates that, in leading order, there are three pairs of fermions with degenerate masses $\tilde{M}_i$.

To incorporate the small contributions from $\mu$ and $\mu'$, we compute the perturbative corrections to both the eigenvalues and eigenvectors of $\mathcal{M}^{(0)}_N$. Our aim is to determine a unitary matrix $Q$ and a diagonal matrix $D=\diag (M_1, \cdots, M_6)$ such that $Q^T \mathcal{M}_N Q = D$. Applying time-independent perturbation theory, the first-order correction to the eigenvalues is given by
\begin{equation}
    \lambda_i^{(1)} = \left< \psi_i^{(0)} \right|\mathcal{M}^{(1)}_N \left| \psi_i^{(0)} \right>,
\end{equation}
where the unperturbed eigenvectors are
 \begin{equation} 
     \begin{cases} 
     \left| \psi_i^{(0)} \right> = \frac{1}{\sqrt{2}} \begin{pmatrix} \vec{u}_{(i)} \\ \vec{v}_{(i)} \end{pmatrix}, & \text{for upper block}; \\ 
     \left| \psi_i^{(0)} \right> = \frac{1}{\sqrt{2}} \begin{pmatrix} -i \vec{u}_{(i)} \\ -i\vec{v}_{(i)} \end{pmatrix}, & \text{for lower block}. 
     \end{cases} 
\end{equation} 
The first-order correction yields identical contributions to both the positive and negative eigenvalues,
\begin{equation}
    \lambda_i^{(1)} = \frac{1}{2} \left( \vec{u}_{(i)}\, ^\dagger \mu' \vec{u}_{(i)} + \vec{v}_{(i)}\, ^\dagger \mu \vec{v}_{(i)} \right).
\end{equation}
However, this uniform shift does not resolve the degeneracy between the paired fermions. Therefore, we must consider second-order corrections to lift the degeneracy
\begin{equation}
      \lambda_i^{(2)} =  \sum_{j \ne i} \frac{\left| \left< \psi_j^{(0)} \right|\mathcal{M}^{(1)}_N \left| \psi_i^{(0)} \right> \right|^2 }{ \tilde{M}_i  - \tilde{M}_j }.
      \label{eq:2ndpert}
\end{equation}
This indeed introduces unequal corrections to the masses within each fermion pair. Additionally, the first-order correction to the eigenvectors is given by
\begin{equation}
    \left| \psi_i^{(1)} \right> = \sum_{j \ne i} \frac{ \left< \psi_j^{(0)} \right|\mathcal{M}^{(1)}_N \left| \psi_i^{(0)} \right> }{ \tilde{M}_i  - \tilde{M}_j } \left| \psi_j^{(0)} \right>,
    \label{eq:1rstwvpert}
\end{equation}
which accounts for the mixing between different unperturbed eigenstates. Matrix elements in the numerators of Eqs.~\eqref{eq:2ndpert} and \eqref{eq:1rstwvpert} can be categorized into four distinct cases:
\begin{equation}
   \left< \psi_j^{(0)} \right|\mathcal{M}^{(1)}_N \left| \psi_i^{(0)} \right> = 
    \begin{cases} 
     \frac{1}{2} \left[ \vec{u}_{(j)}\, ^\dagger \mu' \vec{u}_{(i)} + \vec{v}_{(j)}\, ^\dagger \mu \vec{v}_{(i)} \right], & \text{ for} ~1\le i\neq j \le 3; \\ 
     \frac{i}{2} \left[ -\vec{u}_{(j)}\, ^\dagger \mu' \vec{u}_{(i)} + \vec{v}_{(j)}\, ^\dagger \mu \vec{v}_{(i)} \right], & \text{ for} ~1\le i \le 3, ~ 4\le  j \le 6; \\ 
    \frac{i}{2} \left[ \vec{u}_{(j)}\, ^\dagger \mu' \vec{u}_{(i)} - \vec{v}_{(j)}\, ^\dagger \mu \vec{v}_{(i)} \right], & \text{ for} ~4\le i \le 6, ~ 1\le  j \le 3; \\ 
     \frac{1}{2} \left[ \vec{u}_{(j)}\, ^\dagger \mu' \vec{u}_{(i)} + \vec{v}_{(j)}\, ^\dagger \mu \vec{v}_{(i)} \right], & \text{ for} ~4\le i\neq j \le 6. 
     \end{cases} 
\end{equation}

Finally, the perturbative diagonalization provides the heavy states masses
\begin{equation}
    M_i \approx \tilde{M}_i + \lambda^{(1)}_i + \lambda^{(2)}_i,
\end{equation}
and the unitary matrix $Q$, whose columns are composed of the perturbed eigenvectors,
\begin{equation}
        \left| \psi_i \right> \approx \left| \psi_i^{(0)} \right> + \left| \psi_i^{(1)} \right>.
\end{equation}
The unitary transformation matrix $Q$ includes the rotation from the mass basis to the flavor basis. In particular, this analysis reveals significant mixing between $N$ and $\chi$ states.

%
%

\bibliographystyle{RAN.bst}
\bibliography{biblio}

\begin{thebibliography}{10}
\providecommand{\url}[1]{\texttt{#1}}
\providecommand{\urlprefix}{URL }
\providecommand{\eprint}[2][]{\url{#2}}

\bibitem{Kajita:2016cak}
T.~Kajita, \emph{{Nobel Lecture: Discovery of atmospheric neutrino
  oscillations}},
  \MYhref[journalLinks]{http://dx.doi.org/10.1103/RevModPhys.88.030501}{Rev.
  Mod. Phys.
  }\MYhref[journalLinks]{http://dx.doi.org/10.1103/RevModPhys.88.030501}{\textbf{88}
  (2016) 3 030501}.

\bibitem{McDonald:2016ixn}
A.~B. McDonald, \emph{{Nobel Lecture: The Sudbury Neutrino Observatory:
  Observation of flavor change for solar neutrinos}},
  \MYhref[journalLinks]{http://dx.doi.org/10.1103/RevModPhys.88.030502}{Rev.
  Mod. Phys.
  }\MYhref[journalLinks]{http://dx.doi.org/10.1103/RevModPhys.88.030502}{\textbf{88}
  (2016) 3 030502}.

\bibitem{Bertone:2004pz}
G.~Bertone, D.~Hooper and J.~Silk, \emph{{Particle dark matter: Evidence,
  candidates and constraints}},
  \MYhref[journalLinks]{http://dx.doi.org/10.1016/j.physrep.2004.08.031}{Phys.
  Rept.
  }\MYhref[journalLinks]{http://dx.doi.org/10.1016/j.physrep.2004.08.031}{\textbf{405}
  (2005) 279--390},
  \MYhref[eprintLinks]{http://arxiv.org/abs/hep-ph/0404175}{{\ttfamily
  arXiv:hep-ph/0404175}}.

\bibitem{Arbey:2021gdg}
A.~Arbey and F.~Mahmoudi, \emph{{Dark matter and the early Universe: a
  review}},
  \MYhref[journalLinks]{http://dx.doi.org/10.1016/j.ppnp.2021.103865}{Prog.
  Part. Nucl. Phys.
  }\MYhref[journalLinks]{http://dx.doi.org/10.1016/j.ppnp.2021.103865}{\textbf{119}
  (2021) 103865},
  \MYhref[eprintLinks]{http://arxiv.org/abs/2104.11488}{{\ttfamily
  arXiv:2104.11488 [hep-ph]}}.

\bibitem{Cirelli:2024ssz}
M.~Cirelli, A.~Strumia and J.~Zupan, \emph{{Dark Matter}}  (2024),
  \MYhref[eprintLinks]{http://arxiv.org/abs/2406.01705}{{\ttfamily
  arXiv:2406.01705 [hep-ph]}}.

\bibitem{Mohapatra:1986bd}
R.~N. Mohapatra and J.~W.~F. Valle, \emph{{Neutrino Mass and Baryon Number
  Nonconservation in Superstring Models}},
  \MYhref[journalLinks]{http://dx.doi.org/10.1103/PhysRevD.34.1642}{Phys. Rev.
  D
  }\MYhref[journalLinks]{http://dx.doi.org/10.1103/PhysRevD.34.1642}{\textbf{34}
  (1986) 1642}.

\bibitem{Das:2012ze}
A.~Das and N.~Okada, \emph{{Inverse seesaw neutrino signatures at the LHC and
  ILC}},
  \MYhref[journalLinks]{http://dx.doi.org/10.1103/PhysRevD.88.113001}{Phys.
  Rev. D
  }\MYhref[journalLinks]{http://dx.doi.org/10.1103/PhysRevD.88.113001}{\textbf{88}
  (2013) 113001},
  \MYhref[eprintLinks]{http://arxiv.org/abs/1207.3734}{{\ttfamily
  arXiv:1207.3734 [hep-ph]}}.

\bibitem{Das:2019pua}
A.~Das, S.~Goswami, K.~N. Vishnudath and T.~Nomura, \emph{{Constraining a
  general U(1)$^\prime$ inverse seesaw model from vacuum stability, dark matter
  and collider}},
  \MYhref[journalLinks]{http://dx.doi.org/10.1103/PhysRevD.101.055026}{Phys.
  Rev. D
  }\MYhref[journalLinks]{http://dx.doi.org/10.1103/PhysRevD.101.055026}{\textbf{101}
  (2020) 5 055026},
  \MYhref[eprintLinks]{http://arxiv.org/abs/1905.00201}{{\ttfamily
  arXiv:1905.00201 [hep-ph]}}.

\bibitem{CarcamoHernandez:2019eme}
A.~E. C\'arcamo~Hern\'andez and S.~F. King, \emph{{Littlest Inverse Seesaw
  Model}},
  \MYhref[journalLinks]{http://dx.doi.org/10.1016/j.nuclphysb.2020.114950}{Nucl.
  Phys. B
  }\MYhref[journalLinks]{http://dx.doi.org/10.1016/j.nuclphysb.2020.114950}{\textbf{953}
  (2020) 114950},
  \MYhref[eprintLinks]{http://arxiv.org/abs/1903.02565}{{\ttfamily
  arXiv:1903.02565 [hep-ph]}}.

\bibitem{CentellesChulia:2020dfh}
S.~Centelles~Chuli\'a, R.~Srivastava and A.~Vicente, \emph{{The inverse seesaw
  family: Dirac and Majorana}},
  \MYhref[journalLinks]{http://dx.doi.org/10.1007/JHEP03(2021)248}{JHEP
  }\MYhref[journalLinks]{http://dx.doi.org/10.1007/JHEP03(2021)248}{\textbf{03}
  (2021) 248}, \MYhref[eprintLinks]{http://arxiv.org/abs/2011.06609}{{\ttfamily
  arXiv:2011.06609 [hep-ph]}}.

\bibitem{Law:2012mj}
S.~S.~C. Law and K.~L. McDonald, \emph{{Inverse seesaw and dark matter in
  models with exotic lepton triplets}},
  \MYhref[journalLinks]{http://dx.doi.org/10.1016/j.physletb.2012.06.044}{Phys.
  Lett. B
  }\MYhref[journalLinks]{http://dx.doi.org/10.1016/j.physletb.2012.06.044}{\textbf{713}
  (2012) 490--494},
  \MYhref[eprintLinks]{http://arxiv.org/abs/1204.2529}{{\ttfamily
  arXiv:1204.2529 [hep-ph]}}.

\bibitem{Abada:2014zra}
A.~Abada, G.~Arcadi and M.~Lucente, \emph{{Dark Matter in the minimal Inverse
  Seesaw mechanism}},
  \MYhref[journalLinks]{http://dx.doi.org/10.1088/1475-7516/2014/10/001}{JCAP
  }\MYhref[journalLinks]{http://dx.doi.org/10.1088/1475-7516/2014/10/001}{\textbf{10}
  (2014) 001}, \MYhref[eprintLinks]{http://arxiv.org/abs/1406.6556}{{\ttfamily
  arXiv:1406.6556 [hep-ph]}}.

\bibitem{Mukherjee:2015axj}
A.~Mukherjee and M.~K. Das, \emph{{Neutrino phenomenology and scalar Dark
  Matter with $A_{4}$ flavor symmetry in Inverse and type II seesaw}},
  \MYhref[journalLinks]{http://dx.doi.org/10.1016/j.nuclphysb.2016.10.008}{Nucl.
  Phys. B
  }\MYhref[journalLinks]{http://dx.doi.org/10.1016/j.nuclphysb.2016.10.008}{\textbf{913}
  (2016) 643--663},
  \MYhref[eprintLinks]{http://arxiv.org/abs/1512.02384}{{\ttfamily
  arXiv:1512.02384 [hep-ph]}}.

\bibitem{Abdallah:2019svm}
W.~Abdallah, S.~Choubey and S.~Khan, \emph{{FIMP dark matter candidate(s) in a
  $B-L$ model with inverse seesaw mechanism}},
  \MYhref[journalLinks]{http://dx.doi.org/10.1007/JHEP06(2019)095}{JHEP
  }\MYhref[journalLinks]{http://dx.doi.org/10.1007/JHEP06(2019)095}{\textbf{06}
  (2019) 095}, \MYhref[eprintLinks]{http://arxiv.org/abs/1904.10015}{{\ttfamily
  arXiv:1904.10015 [hep-ph]}}.

\bibitem{Pongkitivanichkul:2019cvm}
C.~Pongkitivanichkul, N.~Thongyoi and P.~Uttayarat, \emph{{Inverse seesaw
  mechanism and portal dark matter}},
  \MYhref[journalLinks]{http://dx.doi.org/10.1103/PhysRevD.100.035034}{Phys.
  Rev. D
  }\MYhref[journalLinks]{http://dx.doi.org/10.1103/PhysRevD.100.035034}{\textbf{100}
  (2019) 3 035034},
  \MYhref[eprintLinks]{http://arxiv.org/abs/1905.13224}{{\ttfamily
  arXiv:1905.13224 [hep-ph]}}.

\bibitem{Mandal:2019oth}
S.~Mandal, N.~Rojas, R.~Srivastava and J.~W.~F. Valle, \emph{{Dark matter as
  the origin of neutrino mass in the inverse seesaw mechanism}},
  \MYhref[journalLinks]{http://dx.doi.org/10.1016/j.physletb.2021.136609}{Phys.
  Lett. B
  }\MYhref[journalLinks]{http://dx.doi.org/10.1016/j.physletb.2021.136609}{\textbf{821}
  (2021) 136609},
  \MYhref[eprintLinks]{http://arxiv.org/abs/1907.07728}{{\ttfamily
  arXiv:1907.07728 [hep-ph]}}.

\bibitem{Gu:2019gzy}
P.-H. Gu, \emph{{Inverse seesaw accompanied by Dirac fermionic dark matter}},
  \MYhref[journalLinks]{http://dx.doi.org/10.1016/j.physletb.2020.135499}{Phys.
  Lett. B
  }\MYhref[journalLinks]{http://dx.doi.org/10.1016/j.physletb.2020.135499}{\textbf{806}
  (2020) 135499},
  \MYhref[eprintLinks]{http://arxiv.org/abs/1907.11556}{{\ttfamily
  arXiv:1907.11556 [hep-ph]}}.

\bibitem{Fernandez-Martinez:2021ypo}
E.~Fernandez-Martinez, M.~Pierre, E.~Pinsard and S.~Rosauro-Alcaraz,
  \emph{{Inverse Seesaw, dark matter and the Hubble tension}},
  \MYhref[journalLinks]{http://dx.doi.org/10.1140/epjc/s10052-021-09760-y}{Eur.
  Phys. J. C
  }\MYhref[journalLinks]{http://dx.doi.org/10.1140/epjc/s10052-021-09760-y}{\textbf{81}
  (2021) 10 954},
  \MYhref[eprintLinks]{http://arxiv.org/abs/2106.05298}{{\ttfamily
  arXiv:2106.05298 [hep-ph]}}.

\bibitem{Abada:2021yot}
A.~Abada et~al., \emph{{Gauged inverse seesaw from dark matter}},
  \MYhref[journalLinks]{http://dx.doi.org/10.1140/epjc/s10052-021-09535-5}{Eur.
  Phys. J. C
  }\MYhref[journalLinks]{http://dx.doi.org/10.1140/epjc/s10052-021-09535-5}{\textbf{81}
  (2021) 8 758},
  \MYhref[eprintLinks]{http://arxiv.org/abs/2107.02803}{{\ttfamily
  arXiv:2107.02803 [hep-ph]}}.

\bibitem{Zhang:2021olk}
X.~Zhang and S.~Zhou, \emph{{Inverse seesaw model with a modular $S_4$
  symmetry: lepton flavor mixing and warm dark~matter}},
  \MYhref[journalLinks]{http://dx.doi.org/10.1088/1475-7516/2021/09/043}{JCAP
  }\MYhref[journalLinks]{http://dx.doi.org/10.1088/1475-7516/2021/09/043}{\textbf{09}
  (2021) 043}, \MYhref[eprintLinks]{http://arxiv.org/abs/2106.03433}{{\ttfamily
  arXiv:2106.03433 [hep-ph]}}.

\bibitem{Gogoi:2023jzl}
J.~Gogoi, L.~Sarma and M.~K. Das, \emph{{Leptogenesis and dark matter in
  minimal inverse seesaw using $A_4$ modular symmetry}},
  \MYhref[journalLinks]{http://dx.doi.org/10.1140/epjc/s10052-024-13029-5}{Eur.
  Phys. J. C
  }\MYhref[journalLinks]{http://dx.doi.org/10.1140/epjc/s10052-024-13029-5}{\textbf{84}
  (2024) 7 689},
  \MYhref[eprintLinks]{http://arxiv.org/abs/2311.09883}{{\ttfamily
  arXiv:2311.09883 [hep-ph]}}.

\bibitem{Zeldovich:1965gev}
Y.~b. Zeldovich, \emph{{Survey of Modern Cosmology}},
  \MYhref[journalLinks]{http://dx.doi.org/10.1016/b978-1-4831-9921-4.50011-9}{Adv.
  Astron. Astrophys.
  }\MYhref[journalLinks]{http://dx.doi.org/10.1016/b978-1-4831-9921-4.50011-9}{\textbf{3}
  (1965) 241--379}.

\bibitem{Scherrer:1985zt}
R.~J. Scherrer and M.~S. Turner, \emph{{On the Relic, Cosmic Abundance of
  Stable Weakly Interacting Massive Particles}},
  \MYhref[journalLinks]{http://dx.doi.org/10.1103/PhysRevD.33.1585}{Phys. Rev.
  D
  }\MYhref[journalLinks]{http://dx.doi.org/10.1103/PhysRevD.33.1585}{\textbf{33}
  (1986) 1585}, [Erratum: Phys.Rev.D 34, 3263 (1986)].

\bibitem{Gondolo:1990dk}
P.~Gondolo and G.~Gelmini, \emph{{Cosmic abundances of stable particles:
  Improved analysis}},
  \MYhref[journalLinks]{http://dx.doi.org/10.1016/0550-3213(91)90438-4}{Nucl.
  Phys. B
  }\MYhref[journalLinks]{http://dx.doi.org/10.1016/0550-3213(91)90438-4}{\textbf{360}
  (1991) 145--179}.

\bibitem{Steigman:2012nb}
G.~Steigman, B.~Dasgupta and J.~F. Beacom, \emph{{Precise Relic WIMP Abundance
  and its Impact on Searches for Dark Matter Annihilation}},
  \MYhref[journalLinks]{http://dx.doi.org/10.1103/PhysRevD.86.023506}{Phys.
  Rev. D
  }\MYhref[journalLinks]{http://dx.doi.org/10.1103/PhysRevD.86.023506}{\textbf{86}
  (2012) 023506},
  \MYhref[eprintLinks]{http://arxiv.org/abs/1204.3622}{{\ttfamily
  arXiv:1204.3622 [hep-ph]}}.

\bibitem{Schumann:2019eaa}
M.~Schumann, \emph{{Direct Detection of WIMP Dark Matter: Concepts and
  Status}},
  \MYhref[journalLinks]{http://dx.doi.org/10.1088/1361-6471/ab2ea5}{J. Phys. G
  }\MYhref[journalLinks]{http://dx.doi.org/10.1088/1361-6471/ab2ea5}{\textbf{46}
  (2019) 10 103003},
  \MYhref[eprintLinks]{http://arxiv.org/abs/1903.03026}{{\ttfamily
  arXiv:1903.03026 [astro-ph.CO]}}.

\bibitem{Frumkin:2022ror}
R.~Frumkin, E.~Kuflik, I.~Lavie and T.~Silverwater, \emph{{Roadmap to Thermal
  Dark Matter beyond the Weakly Interacting Dark Matter Unitarity Bound}},
  \MYhref[journalLinks]{http://dx.doi.org/10.1103/PhysRevLett.130.171001}{Phys.
  Rev. Lett.
  }\MYhref[journalLinks]{http://dx.doi.org/10.1103/PhysRevLett.130.171001}{\textbf{130}
  (2023) 17 171001},
  \MYhref[eprintLinks]{http://arxiv.org/abs/2207.01635}{{\ttfamily
  arXiv:2207.01635 [hep-ph]}}.

\bibitem{delaTorre:2023nfk}
P.~de~la Torre, M.~Guti\'errez and M.~Masip, \emph{{Monochromatic neutrinos
  from dark matter through the Higgs portal}},
  \MYhref[journalLinks]{http://dx.doi.org/10.1088/1475-7516/2023/11/068}{JCAP
  }\MYhref[journalLinks]{http://dx.doi.org/10.1088/1475-7516/2023/11/068}{\textbf{11}
  (2023) 068}, \MYhref[eprintLinks]{http://arxiv.org/abs/2309.00374}{{\ttfamily
  arXiv:2309.00374 [hep-ph]}}.

\bibitem{Hall:2009bx}
L.~J. Hall, K.~Jedamzik, J.~March-Russell and S.~M. West, \emph{{Freeze-In
  Production of FIMP Dark Matter}},
  \MYhref[journalLinks]{http://dx.doi.org/10.1007/JHEP03(2010)080}{JHEP
  }\MYhref[journalLinks]{http://dx.doi.org/10.1007/JHEP03(2010)080}{\textbf{03}
  (2010) 080}, \MYhref[eprintLinks]{http://arxiv.org/abs/0911.1120}{{\ttfamily
  arXiv:0911.1120 [hep-ph]}}.

\bibitem{Elahi:2014fsa}
F.~Elahi, C.~Kolda and J.~Unwin, \emph{{UltraViolet Freeze-in}},
  \MYhref[journalLinks]{http://dx.doi.org/10.1007/JHEP03(2015)048}{JHEP
  }\MYhref[journalLinks]{http://dx.doi.org/10.1007/JHEP03(2015)048}{\textbf{03}
  (2015) 048}, \MYhref[eprintLinks]{http://arxiv.org/abs/1410.6157}{{\ttfamily
  arXiv:1410.6157 [hep-ph]}}.

\bibitem{Biswas:2016bfo}
A.~Biswas and A.~Gupta, \emph{{Freeze-in Production of Sterile Neutrino Dark
  Matter in U(1)$_{\rm B-L}$ Model}},
  \MYhref[journalLinks]{http://dx.doi.org/10.1088/1475-7516/2016/09/044}{JCAP
  }\MYhref[journalLinks]{http://dx.doi.org/10.1088/1475-7516/2016/09/044}{\textbf{09}
  (2016) 044}, [Addendum: JCAP 05, A01 (2017)],
  \MYhref[eprintLinks]{http://arxiv.org/abs/1607.01469}{{\ttfamily
  arXiv:1607.01469 [hep-ph]}}.

\bibitem{Bernal:2017kxu}
N.~Bernal et~al., \emph{{The Dawn of FIMP Dark Matter: A Review of Models and
  Constraints}},
  \MYhref[journalLinks]{http://dx.doi.org/10.1142/S0217751X1730023X}{Int. J.
  Mod. Phys. A
  }\MYhref[journalLinks]{http://dx.doi.org/10.1142/S0217751X1730023X}{\textbf{32}
  (2017) 27 1730023},
  \MYhref[eprintLinks]{http://arxiv.org/abs/1706.07442}{{\ttfamily
  arXiv:1706.07442 [hep-ph]}}.

\bibitem{Chianese:2018dsz}
M.~Chianese and S.~F. King, \emph{{The Dark Side of the Littlest Seesaw:
  freeze-in, the two right-handed neutrino portal and leptogenesis-friendly
  fimpzillas}},
  \MYhref[journalLinks]{http://dx.doi.org/10.1088/1475-7516/2018/09/027}{JCAP
  }\MYhref[journalLinks]{http://dx.doi.org/10.1088/1475-7516/2018/09/027}{\textbf{09}
  (2018) 027}, \MYhref[eprintLinks]{http://arxiv.org/abs/1806.10606}{{\ttfamily
  arXiv:1806.10606 [hep-ph]}}.

\bibitem{Becker:2018rve}
M.~Becker, \emph{{Dark Matter from Freeze-In via the Neutrino Portal}},
  \MYhref[journalLinks]{http://dx.doi.org/10.1140/epjc/s10052-019-7095-7}{Eur.
  Phys. J. C
  }\MYhref[journalLinks]{http://dx.doi.org/10.1140/epjc/s10052-019-7095-7}{\textbf{79}
  (2019) 7 611},
  \MYhref[eprintLinks]{http://arxiv.org/abs/1806.08579}{{\ttfamily
  arXiv:1806.08579 [hep-ph]}}.

\bibitem{Chianese:2019epo}
M.~Chianese, B.~Fu and S.~F. King, \emph{{Minimal Seesaw extension for Neutrino
  Mass and Mixing, Leptogenesis and Dark Matter: FIMPzillas through the
  Right-Handed Neutrino Portal}},
  \MYhref[journalLinks]{http://dx.doi.org/10.1088/1475-7516/2020/03/030}{JCAP
  }\MYhref[journalLinks]{http://dx.doi.org/10.1088/1475-7516/2020/03/030}{\textbf{03}
  (2020) 030}, \MYhref[eprintLinks]{http://arxiv.org/abs/1910.12916}{{\ttfamily
  arXiv:1910.12916 [hep-ph]}}.

\bibitem{Allahverdi:2019jsc}
R.~Allahverdi and J.~K. Osi\'nski, \emph{{Freeze-in Production of Dark Matter
  Prior to Early Matter Domination}},
  \MYhref[journalLinks]{http://dx.doi.org/10.1103/PhysRevD.101.063503}{Phys.
  Rev. D
  }\MYhref[journalLinks]{http://dx.doi.org/10.1103/PhysRevD.101.063503}{\textbf{101}
  (2020) 6 063503},
  \MYhref[eprintLinks]{http://arxiv.org/abs/1909.01457}{{\ttfamily
  arXiv:1909.01457 [hep-ph]}}.

\bibitem{Cosme:2020mck}
C.~Cosme et~al., \emph{{Neutrino portal to FIMP dark matter with an early
  matter era}},
  \MYhref[journalLinks]{http://dx.doi.org/10.1007/JHEP03(2021)026}{JHEP
  }\MYhref[journalLinks]{http://dx.doi.org/10.1007/JHEP03(2021)026}{\textbf{03}
  (2021) 026}, \MYhref[eprintLinks]{http://arxiv.org/abs/2003.01723}{{\ttfamily
  arXiv:2003.01723 [hep-ph]}}.

\bibitem{Das:2021nqj}
A.~Das, S.~Goswami, V.~K.~N. and T.~K. Poddar, \emph{{Freeze-in production of
  sterile neutrino dark matter in a gauged U(1)' model with inverse seesaw}},
  \MYhref[journalLinks]{http://dx.doi.org/10.1016/j.nuclphysb.2024.116568}{Nucl.
  Phys. B
  }\MYhref[journalLinks]{http://dx.doi.org/10.1016/j.nuclphysb.2024.116568}{\textbf{1004}
  (2024) 116568},
  \MYhref[eprintLinks]{http://arxiv.org/abs/2104.13986}{{\ttfamily
  arXiv:2104.13986 [hep-ph]}}.

\bibitem{Yaguna:2023kyu}
C.~E. Yaguna and O.~Zapata, \emph{{Minimal model of fermion FIMP dark matter}},
  \MYhref[journalLinks]{http://dx.doi.org/10.1103/PhysRevD.109.015002}{Phys.
  Rev. D
  }\MYhref[journalLinks]{http://dx.doi.org/10.1103/PhysRevD.109.015002}{\textbf{109}
  (2024) 1 015002},
  \MYhref[eprintLinks]{http://arxiv.org/abs/2308.05249}{{\ttfamily
  arXiv:2308.05249 [hep-ph]}}.

\bibitem{Abada:2023mib}
A.~Abada et~al., \emph{{Thermal effects in freeze-in neutrino dark mater
  production}},
  \MYhref[journalLinks]{http://dx.doi.org/10.1007/JHEP11(2023)180}{JHEP
  }\MYhref[journalLinks]{http://dx.doi.org/10.1007/JHEP11(2023)180}{\textbf{11}
  (2023) 180}, \MYhref[eprintLinks]{http://arxiv.org/abs/2308.01341}{{\ttfamily
  arXiv:2308.01341 [hep-ph]}}.

\bibitem{Xu:2023xva}
X.-J. Xu, S.~Zhou and J.~Zhu, \emph{{The $\nu_{R}$-philic scalar dark matter}},
  \MYhref[journalLinks]{http://dx.doi.org/10.1088/1475-7516/2024/04/012}{JCAP
  }\MYhref[journalLinks]{http://dx.doi.org/10.1088/1475-7516/2024/04/012}{\textbf{04}
  (2024) 012}, \MYhref[eprintLinks]{http://arxiv.org/abs/2310.16346}{{\ttfamily
  arXiv:2310.16346 [hep-ph]}}.

\bibitem{Schmaltz:2017oov}
M.~Schmaltz and N.~Weiner, \emph{{A Portalino to the Dark Sector}},
  \MYhref[journalLinks]{http://dx.doi.org/10.1007/JHEP02(2019)105}{JHEP
  }\MYhref[journalLinks]{http://dx.doi.org/10.1007/JHEP02(2019)105}{\textbf{02}
  (2019) 105}, \MYhref[eprintLinks]{http://arxiv.org/abs/1709.09164}{{\ttfamily
  arXiv:1709.09164 [hep-ph]}}.

\bibitem{Yaguna:2019cvp}
C.~E. Yaguna and O.~Zapata, \emph{{Multi-component scalar dark matter from a
  $Z_N$ symmetry: a systematic analysis}},
  \MYhref[journalLinks]{http://dx.doi.org/10.1007/JHEP03(2020)109}{JHEP
  }\MYhref[journalLinks]{http://dx.doi.org/10.1007/JHEP03(2020)109}{\textbf{03}
  (2020) 109}, \MYhref[eprintLinks]{http://arxiv.org/abs/1911.05515}{{\ttfamily
  arXiv:1911.05515 [hep-ph]}}.

\bibitem{Yaguna:2021vhb}
C.~E. Yaguna and O.~Zapata, \emph{{Two-component scalar dark matter in Z$_{2n}$
  scenarios}},
  \MYhref[journalLinks]{http://dx.doi.org/10.1007/JHEP10(2021)185}{JHEP
  }\MYhref[journalLinks]{http://dx.doi.org/10.1007/JHEP10(2021)185}{\textbf{10}
  (2021) 185}, \MYhref[eprintLinks]{http://arxiv.org/abs/2106.11889}{{\ttfamily
  arXiv:2106.11889 [hep-ph]}}.

\bibitem{Hepburn:2022pin}
D.~Hepburn and S.~M. West, \emph{{Dark matter and neutrino masses in a
  Portalino-like model}},
  \MYhref[journalLinks]{http://dx.doi.org/10.1140/epjc/s10052-023-11493-z}{Eur.
  Phys. J. C
  }\MYhref[journalLinks]{http://dx.doi.org/10.1140/epjc/s10052-023-11493-z}{\textbf{83}
  (2023) 5 405},
  \MYhref[eprintLinks]{http://arxiv.org/abs/2208.02698}{{\ttfamily
  arXiv:2208.02698 [hep-ph]}}.

\bibitem{Liu:2023kil}
A.~Liu, Z.-L. Han, Y.~Jin and H.~Li, \emph{{Sterile neutrino portal dark matter
  with $Z_3$ symmetry}},
  \MYhref[journalLinks]{http://dx.doi.org/10.1103/PhysRevD.108.075021}{Phys.
  Rev. D
  }\MYhref[journalLinks]{http://dx.doi.org/10.1103/PhysRevD.108.075021}{\textbf{108}
  (2023) 7 075021},
  \MYhref[eprintLinks]{http://arxiv.org/abs/2306.14091}{{\ttfamily
  arXiv:2306.14091 [hep-ph]}}.

\bibitem{Kim:2024cwp}
J.~Kim, S.-S. Kim, H.~M. Lee and R.~Padhan, \emph{{Small Neutrino Masses from a
  Decoupled Singlet Scalar Field}}  (2024),
  \MYhref[eprintLinks]{http://arxiv.org/abs/2407.13595}{{\ttfamily
  arXiv:2407.13595 [hep-ph]}}.

\bibitem{tHooft:1979rat}
G.~'t~Hooft, \emph{{Naturalness, chiral symmetry, and spontaneous chiral
  symmetry breaking}},
  \MYhref[journalLinks]{http://dx.doi.org/10.1007/978-1-4684-7571-5_9}{NATO
  Sci. Ser. B
  }\MYhref[journalLinks]{http://dx.doi.org/10.1007/978-1-4684-7571-5_9}{\textbf{59}
  (1980) 135--157}.

\bibitem{Weinberg:1979sa}
S.~Weinberg, \emph{{Baryon and Lepton Nonconserving Processes}},
  \MYhref[journalLinks]{http://dx.doi.org/10.1103/PhysRevLett.43.1566}{Phys.
  Rev. Lett.
  }\MYhref[journalLinks]{http://dx.doi.org/10.1103/PhysRevLett.43.1566}{\textbf{43}
  (1979) 1566--1570}.

\bibitem{Robens:2015gla}
T.~Robens and T.~Stefaniak, \emph{{Status of the Higgs Singlet Extension of the
  Standard Model after LHC Run 1}},
  \MYhref[journalLinks]{http://dx.doi.org/10.1140/epjc/s10052-015-3323-y}{Eur.
  Phys. J. C
  }\MYhref[journalLinks]{http://dx.doi.org/10.1140/epjc/s10052-015-3323-y}{\textbf{75}
  (2015) 104}, \MYhref[eprintLinks]{http://arxiv.org/abs/1501.02234}{{\ttfamily
  arXiv:1501.02234 [hep-ph]}}.

\bibitem{Dupuis:2016fda}
G.~Dupuis, \emph{{Collider Constraints and Prospects of a Scalar Singlet
  Extension to Higgs Portal Dark Matter}},
  \MYhref[journalLinks]{http://dx.doi.org/10.1007/JHEP07(2016)008}{JHEP
  }\MYhref[journalLinks]{http://dx.doi.org/10.1007/JHEP07(2016)008}{\textbf{07}
  (2016) 008}, \MYhref[eprintLinks]{http://arxiv.org/abs/1604.04552}{{\ttfamily
  arXiv:1604.04552 [hep-ph]}}.

\bibitem{Bechtle:2020uwn}
P.~Bechtle et~al., \emph{{HiggsSignals-2: Probing new physics with precision
  Higgs measurements in the LHC 13 TeV era}},
  \MYhref[journalLinks]{http://dx.doi.org/10.1140/epjc/s10052-021-08942-y}{Eur.
  Phys. J. C
  }\MYhref[journalLinks]{http://dx.doi.org/10.1140/epjc/s10052-021-08942-y}{\textbf{81}
  (2021) 2 145},
  \MYhref[eprintLinks]{http://arxiv.org/abs/2012.09197}{{\ttfamily
  arXiv:2012.09197 [hep-ph]}}.

\bibitem{Lane:2024vur}
S.~D. Lane, I.~M. Lewis and M.~Sullivan, \emph{{Resonant multiscalar production
  in the generic complex singlet model in the multi-TeV region}},
  \MYhref[journalLinks]{http://dx.doi.org/10.1103/PhysRevD.110.055017}{Phys.
  Rev. D
  }\MYhref[journalLinks]{http://dx.doi.org/10.1103/PhysRevD.110.055017}{\textbf{110}
  (2024) 5 055017},
  \MYhref[eprintLinks]{http://arxiv.org/abs/2403.18003}{{\ttfamily
  arXiv:2403.18003 [hep-ph]}}.

\bibitem{Ishimori:2010au}
H.~Ishimori et~al., \emph{{Non-Abelian Discrete Symmetries in Particle
  Physics}}, \MYhref[journalLinks]{http://dx.doi.org/10.1143/PTPS.183.1}{Prog.
  Theor. Phys. Suppl.
  }\MYhref[journalLinks]{http://dx.doi.org/10.1143/PTPS.183.1}{\textbf{183}
  (2010) 1--163},
  \MYhref[eprintLinks]{http://arxiv.org/abs/1003.3552}{{\ttfamily
  arXiv:1003.3552 [hep-th]}}.

\bibitem{Novichkov:2019sqv}
P.~P. Novichkov, J.~T. Penedo, S.~T. Petcov and A.~V. Titov, \emph{{Generalised
  CP Symmetry in Modular-Invariant Models of Flavour}},
  \MYhref[journalLinks]{http://dx.doi.org/10.1007/JHEP07(2019)165}{JHEP
  }\MYhref[journalLinks]{http://dx.doi.org/10.1007/JHEP07(2019)165}{\textbf{07}
  (2019) 165}, \MYhref[eprintLinks]{http://arxiv.org/abs/1905.11970}{{\ttfamily
  arXiv:1905.11970 [hep-ph]}}.

\bibitem{FlavorPy}
A.~Baur, \emph{{FlavorPy}} (2024),
  \urlprefix\url{https://doi.org/10.5281/zenodo.11060597}.

\bibitem{Novichkov:2020eep}
P.~P. Novichkov, J.~T. Penedo and S.~T. Petcov, \emph{{Double cover of modular
  $S_4$ for flavour model building}},
  \MYhref[journalLinks]{http://dx.doi.org/10.1016/j.nuclphysb.2020.115301}{Nucl.
  Phys. B
  }\MYhref[journalLinks]{http://dx.doi.org/10.1016/j.nuclphysb.2020.115301}{\textbf{963}
  (2021) 115301},
  \MYhref[eprintLinks]{http://arxiv.org/abs/2006.03058}{{\ttfamily
  arXiv:2006.03058 [hep-ph]}}.

\bibitem{Esteban:2020cvm}
I.~Esteban et~al., \emph{{The fate of hints: updated global analysis of
  three-flavor neutrino oscillations}},
  \MYhref[journalLinks]{http://dx.doi.org/10.1007/JHEP09(2020)178}{JHEP
  }\MYhref[journalLinks]{http://dx.doi.org/10.1007/JHEP09(2020)178}{\textbf{09}
  (2020) 178}, \MYhref[eprintLinks]{http://arxiv.org/abs/2007.14792}{{\ttfamily
  arXiv:2007.14792 [hep-ph]}}.

\bibitem{McDonald:2001vt}
J.~McDonald, \emph{{Thermally generated gauge singlet scalars as
  selfinteracting dark matter}},
  \MYhref[journalLinks]{http://dx.doi.org/10.1103/PhysRevLett.88.091304}{Phys.
  Rev. Lett.
  }\MYhref[journalLinks]{http://dx.doi.org/10.1103/PhysRevLett.88.091304}{\textbf{88}
  (2002) 091304},
  \MYhref[eprintLinks]{http://arxiv.org/abs/hep-ph/0106249}{{\ttfamily
  arXiv:hep-ph/0106249}}.

\bibitem{Belanger:2018ccd}
G.~B\'elanger et~al., \emph{{micrOMEGAs5.0 : Freeze-in}},
  \MYhref[journalLinks]{http://dx.doi.org/10.1016/j.cpc.2018.04.027}{Comput.
  Phys. Commun.
  }\MYhref[journalLinks]{http://dx.doi.org/10.1016/j.cpc.2018.04.027}{\textbf{231}
  (2018) 173--186},
  \MYhref[eprintLinks]{http://arxiv.org/abs/1801.03509}{{\ttfamily
  arXiv:1801.03509 [hep-ph]}}.

\bibitem{Planck:2018vyg}
N.~Aghanim et~al. (Planck), \emph{{Planck 2018 results. VI. Cosmological
  parameters}},
  \MYhref[journalLinks]{http://dx.doi.org/10.1051/0004-6361/201833910}{Astron.
  Astrophys.
  }\MYhref[journalLinks]{http://dx.doi.org/10.1051/0004-6361/201833910}{\textbf{641}
  (2020) A6}, [Erratum: Astron.Astrophys. 652, C4 (2021)],
  \MYhref[eprintLinks]{http://arxiv.org/abs/1807.06209}{{\ttfamily
  arXiv:1807.06209 [astro-ph.CO]}}.

\bibitem{Zeldovich:1974uw}
Y.~B. Zeldovich, I.~Y. Kobzarev and L.~B. Okun, \emph{{Cosmological
  Consequences of the Spontaneous Breakdown of Discrete Symmetry}}, Zh. Eksp.
  Teor. Fiz. \textbf{67} (1974) 3--11.

\bibitem{Kibble:1976sj}
T.~W.~B. Kibble, \emph{{Topology of Cosmic Domains and Strings}},
  \MYhref[journalLinks]{http://dx.doi.org/10.1088/0305-4470/9/8/029}{J. Phys. A
  }\MYhref[journalLinks]{http://dx.doi.org/10.1088/0305-4470/9/8/029}{\textbf{9}
  (1976) 1387--1398}.

\bibitem{Vilenkin:1981zs}
A.~Vilenkin, \emph{{Gravitational Field of Vacuum Domain Walls and Strings}},
  \MYhref[journalLinks]{http://dx.doi.org/10.1103/PhysRevD.23.852}{Phys. Rev. D
  }\MYhref[journalLinks]{http://dx.doi.org/10.1103/PhysRevD.23.852}{\textbf{23}
  (1981) 852--857}.

\bibitem{Coulson:1995nv}
D.~Coulson, Z.~Lalak and B.~A. Ovrut, \emph{{Biased domain walls}},
  \MYhref[journalLinks]{http://dx.doi.org/10.1103/PhysRevD.53.4237}{Phys. Rev.
  D
  }\MYhref[journalLinks]{http://dx.doi.org/10.1103/PhysRevD.53.4237}{\textbf{53}
  (1996) 4237--4246}.

\bibitem{Krajewski:2021jje}
T.~Krajewski, J.~H. Kwapisz, Z.~Lalak and M.~Lewicki, \emph{{Stability of
  domain walls in models with asymmetric potentials}},
  \MYhref[journalLinks]{http://dx.doi.org/10.1103/PhysRevD.104.123522}{Phys.
  Rev. D
  }\MYhref[journalLinks]{http://dx.doi.org/10.1103/PhysRevD.104.123522}{\textbf{104}
  (2021) 12 123522},
  \MYhref[eprintLinks]{http://arxiv.org/abs/2103.03225}{{\ttfamily
  arXiv:2103.03225 [astro-ph.CO]}}.

\bibitem{Gelmini:1988sf}
G.~B. Gelmini, M.~Gleiser and E.~W. Kolb, \emph{{Cosmology of Biased Discrete
  Symmetry Breaking}},
  \MYhref[journalLinks]{http://dx.doi.org/10.1103/PhysRevD.39.1558}{Phys. Rev.
  D
  }\MYhref[journalLinks]{http://dx.doi.org/10.1103/PhysRevD.39.1558}{\textbf{39}
  (1989) 1558}.

\bibitem{Kamionkowski:1992mf}
M.~Kamionkowski and J.~March-Russell, \emph{{Planck scale physics and the
  Peccei-Quinn mechanism}},
  \MYhref[journalLinks]{http://dx.doi.org/10.1016/0370-2693(92)90492-M}{Phys.
  Lett. B
  }\MYhref[journalLinks]{http://dx.doi.org/10.1016/0370-2693(92)90492-M}{\textbf{282}
  (1992) 137--141},
  \MYhref[eprintLinks]{http://arxiv.org/abs/hep-th/9202003}{{\ttfamily
  arXiv:hep-th/9202003}}.

\bibitem{Holman:1992us}
R.~Holman et~al., \emph{{Solutions to the strong CP problem in a world with
  gravity}},
  \MYhref[journalLinks]{http://dx.doi.org/10.1016/0370-2693(92)90491-L}{Phys.
  Lett. B
  }\MYhref[journalLinks]{http://dx.doi.org/10.1016/0370-2693(92)90491-L}{\textbf{282}
  (1992) 132--136},
  \MYhref[eprintLinks]{http://arxiv.org/abs/hep-ph/9203206}{{\ttfamily
  arXiv:hep-ph/9203206}}.

\bibitem{Larsson:1996sp}
S.~E. Larsson, S.~Sarkar and P.~L. White, \emph{{Evading the cosmological
  domain wall problem}},
  \MYhref[journalLinks]{http://dx.doi.org/10.1103/PhysRevD.55.5129}{Phys. Rev.
  D
  }\MYhref[journalLinks]{http://dx.doi.org/10.1103/PhysRevD.55.5129}{\textbf{55}
  (1997) 5129--5135},
  \MYhref[eprintLinks]{http://arxiv.org/abs/hep-ph/9608319}{{\ttfamily
  arXiv:hep-ph/9608319}}.

\bibitem{Li:2025gld}
Y.-J. Li, J.~Liu and Z.-K. Guo, \emph{{Dynamics of $Z_N$ domain walls with bias
  directions}}  (2025),
  \MYhref[eprintLinks]{http://arxiv.org/abs/2502.13644}{{\ttfamily
  arXiv:2502.13644 [astro-ph.CO]}}.

\bibitem{Borboruah:2024lli}
Z.~A. Borboruah, D.~Borah, L.~Malhotra and U.~Patel, \emph{{Minimal Dirac
  seesaw dark matter}}  (2024),
  \MYhref[eprintLinks]{http://arxiv.org/abs/2412.12267}{{\ttfamily
  arXiv:2412.12267 [hep-ph]}}.

\end{thebibliography}

\end{document}